\documentclass[reprint,amssymb,amsmath,aps,unsortedaddress,superscriptaddress, showpacs, footinbib,prb]{revtex4-1}
\usepackage{graphicx}
\usepackage{multirow}
\usepackage{color}
\usepackage[colorlinks,breaklinks,bookmarks=true,citecolor=blue,linkcolor=red,urlcolor=blue]{hyperref}
\newcommand{\crbpo}{Cr$_2[$BP$_3$O$_{12}]$}
\newcommand{\jici}{\ensuremath{J_{\rm ic1}}}
\newcommand{\jicii}{\ensuremath{J_{\rm ic2}}}
\newcommand{\kv}{\mathbf k}
\newcommand{\eff}{\text{eff}}

\begin{document}

\title{Structure and magnetism of Cr$_2[$BP$_3$O$_{12}]$:\\ Towards the
``quantum--classical'' crossover in a spin-3/2 alternating chain }

\author{O.~Janson}
\email{janson@cpfs.mpg.de}
\affiliation{Max Planck Institute for Chemical Physics of Solids, 01187 Dresden, Germany}

\author{S.~Chen}
\affiliation{Max Planck Institute for Chemical Physics of Solids, 01187 Dresden, Germany}

\author{A.~A.~Tsirlin}
\email{altsirlin@gmail.com}
\affiliation{Max Planck Institute for Chemical Physics of Solids, 01187 Dresden, Germany}
\affiliation{National Institute of Chemical Physics and Biophysics, 12618 Tallinn, Estonia}

\author{S.~Hoffmann}
\affiliation{Max Planck Institute for Chemical Physics of Solids, 01187 Dresden, Germany}

\author{J.~Sichelschmidt}
\affiliation{Max Planck Institute for Chemical Physics of Solids, 01187 Dresden, Germany}

\author{Q.~Huang}
\affiliation{NIST Center for Neutron Research,
Gaithersburg, MD 20899, USA}

\author{Z.-J.~Zhang}
\affiliation{Key Laboratory of Transparent Opto-Functional Inorganic Materials
of Chinese Academy of Sciences, Shanghai 200050, PR China}

\author{M.-B.~Tang}
\affiliation{Key Laboratory of Transparent Opto-Functional Inorganic Materials
of Chinese Academy of Sciences, Shanghai 200050, PR China}

\author{J.-T.~Zhao}
\affiliation{Key Laboratory of Transparent Opto-Functional Inorganic Materials
of Chinese Academy of Sciences, Shanghai 200050, PR China}

\author{R.~Kniep}
\affiliation{Max Planck Institute for Chemical Physics of Solids, 01187 Dresden, Germany}

\author{H.~Rosner}
\email{rosner@cpfs.mpg.de}
\affiliation{Max Planck Institute for Chemical Physics of Solids, 01187 Dresden, Germany}

\date{\today}

\begin{abstract}
Magnetic properties of the spin-3/2 Heisenberg system Cr$_2[$BP$_3$O$_{12}]$ are
investigated by magnetic susceptibility $\chi(T)$ measurements, electron spin
resonance, neutron diffraction, and density functional theory (DFT)
calculations, as well as classical and quantum Monte Carlo (MC) simulations.
The broad maximum of $\chi(T)$ at 85\,K and the antiferromagnetic Weiss
temperature of 139\,K indicate low-dimensional magnetic behavior. Below $T_{\rm
N}\!=\!28$\,K, Cr$_2[$BP$_3$O$_{12}]$ is antiferromagnetically  ordered with the
$\mathbf{k}=0$ propagation vector and an ordered moment of 2.5\,$\mu_{\rm
B}$/Cr. DFT calculations, including DFT+$U$ and hybrid functionals, yield a
microscopic model of spin chains with alternating nearest-neighbor couplings
$J_1$ and $J_1'$. The chains are coupled by two nonequivalent interchain
exchanges of similar strength ($\sim$1--2\,K), but different sign
(antiferromagnetic and ferromagnetic). The resulting spin lattice is
quasi-one-dimensional and not frustrated. Quantum MC simulations show excellent
agreement with the experimental data for the parameters $J_1\simeq 50$\,K and
$J_1'/J_1\!\simeq\!0.5$. Therefore, Cr$_2[$BP$_3$O$_{12}]$ is close to the
gapless critical point ($J_1'/J_1=0.41$) of the spin-3/2 bond-alternating
Heisenberg chain. The applicability limits of the classical approximation are
addressed by quantum and classical MC simulations. Implications for a wide
range of low-dimensional $S$\,=\,3/2 materials are discussed.
\end{abstract}


\maketitle

\section{Introduction}
Magnetic properties of transition-metal compounds are generally described with
the Heisenberg spin Hamiltonian that may be augmented by additional terms
responsible for the anisotropy. For example, the following Hamiltonian
\begin{equation}
\label{E-H}
H = \sum_{\langle{}ij\rangle}J\mathbf{S}_i\cdot\mathbf{S}_j -
\sum_iD_i\left(\mathbf{S}_i\cdot{}\hat{z}\right)^2
\end{equation}
accounts for a variety of experimental situations. Here, the
first term describes the isotropic coupling $J$ between spins on sites $i$ and
$j$, the second term is the single-ion anisotropy $D_i$, and $\hat{z}$ is a
unitary vector along $z$, whereas $\mathbf{S}_i$ and $\mathbf{S}_j$ are
quantum-mechanical spin operators.  Although a handful of solvable Heisenberg
models exist,\cite{Bethe_Ansatz, *MG_1, *MG_2, Shastry_Suth} the complex
algebra of the spin operators generally impedes analytical solutions. 

A widely used approximation of a Heisenberg model is its classical
treatment: the original quantum mechanical spin operators $\mathbf{S}_i$
are replaced by real-space vectors $\vec{S}_i$.  This transformation
leads to enormous alleviation of the computational effort, even if the
topology of exchange couplings is very intricate.

A fundamental limitation of classical models is their inability to account for
quantum-mechanical singlets and the ensuing underestimation of the ground state 
(GS) energy. Indeed, for an antiferromagnetic (AF) exchange coupling $J$, the
energy of the quantum-mechanical singlet state \mbox{$-JS(S+1)$} is always
lower than the classical energy $-JS^2$ of the antiparallel spin arrangement.  At
the same time, the relative energy gain of a singlet scales as $S^{-1}$, thus
being maximal for $S\!=\!1/2$ and infinitesimal in the $S\rightarrow\infty$
limit.

Presently, there is an empirical evidence that for $S\!\geq\!2$, classical
models capture the essential physics and correctly reproduce the experimental
magnetic
behavior.\cite{dingle1969,*hutchings1972,dejonge1978,itoh1997,*itoh2002} In
contrast, classical models often fail to predict the correct GS and magnetic
excitation spectrum for the extreme ``quantum'' case of $S\!=\!1/2$.\cite{[{For
example: }][{}]chandra1988,*dagotto1989,[{For example:
}][{}]rousochatzakis2012}  Further on, many $S\!=\!1$ systems, e.g.,\
one-dimensional (1D) Haldane chains,\cite{haldane1983} can not be treated
classically. Therefore, it is crucial to establish the applicability limits for
the classical approximation, in order to distinguish between the ``classical''
cases amenable to a simplified model treatment, and the ``quantum'' cases that
require the complete quantum-mechanical solution of the spin Hamiltonian.

A feasible way towards better understanding of these limits are real material
studies, that allow for a direct comparison between theory and experiment.
Previous work on the quantum-classical crossover rendered the quasi-1D
$S\!=\!3/2$ magnets as the relevant playground. Despite the relatively large
spin of 3/2, the pronounced one-dimensionality may lead to sizable quantum
fluctuations and, thus, to deviations from the classical behavior. For example,
inelastic neutron scattering (INS) studies of Cs$[$V$X_3]$
($X$\,=\,Cl, Br) evidence that at high temperatures these materials are
classical, while lowering the temperature results in a crossover to the
quantum behavior.\cite{CsVCl3_INS_1995, *CsVCl3_CsVBr3_INS_1999} The
INS experiments on the quasi-1D magnet AgCr$[$P$_2$S$_6]$ yielded a sizable
discrepancy between the observed spin-wave velocity and its classical
value, thus indicating strong quantum effects that are present in
$S\!=\!3/2$ chains.\cite{AgCrP2S6_1992, *AgCrP2S6_INS_1993} 

Here, we present a joint experimental and theoretical study of the quasi-1D
$S\!=\!3/2$ system \crbpo. Its crystal structure features magnetic Cr(III)$_2$O$_9$
blocks embedded into a complex borophosphate framework
(Fig.~\ref{F-str}). Although this type of crystal structure could lead to the
simple magnetism of isolated spin dimers, neutron diffraction and magnetic
susceptibility measurements evidence long-range AF ordering at $T_{\rm
N}\!=\!28$\,K that is indicative of sizable interdimer couplings. Using
extensive density-functional theory (DFT) calculations, we evaluate a
microscopic magnetic model for this compound and establish the spin lattice of
weakly coupled bond-alternating Heisenberg chains. This theoretical model is in
good agreement with the neutron scattering and magnetic susceptibility data.

As the microscopic magnetic model of \crbpo\ lacks frustration, its properties
can be simulated with the computationally efficient quantum Monte Carlo (QMC)
techniques. At the same time, the classical model on the same spin lattice can
be treated using classical Monte Carlo (MC) algorithms. By comparing the QMC
and classical MC results, we evaluate the relative impact of quantum as well as
thermal fluctuations on the spin correlations, and in this way address the
crossover between the quantum and classical behavior. 

This paper is organized as follows. The experimental as well as numerical
methods are described in Sec.~\ref{S-method}.  The experimental part in
Sec.~\ref{S-exp} comprises the results of neutron diffraction, electron spin
resonance (ESR), and magnetic susceptibility measurements.  The DFT-based
evaluation of the microscopic magnetic model (Sec.~\ref{S-DFT}) is followed by
the refinement of model parameters by means of QMC simulations and subsequent
fitting to the experiment (Sec.~\ref{S-simul}).  The differences between the
quantum model and its classical counterpart are discussed in
Sec.~\ref{S-quant-class}. We summarize our results and give a short outlook in
Sec.~\ref{S-sum}.

\section{\label{S-method}Methods}
\crbpo\ was prepared by the Pechini-type method.\cite{pechini} 2.172 g chromium
acetate (Alfa Aesar, 23.37(5)\,wt\% chromium content) and 5.7945 g citric acid
(Alfa Aesar 99+\%) were dissolved in 50\,ml water. 5\,ml of glycerol (Sigma
99\%) and a well-ground mixture of 0.3197\,g boric acid (Sigma) and 1.7613\,g
ammonium dihydrogen phosphate (Merck) were added to the dark-green solution.
After slow evaporation of water, a transparent resin was formed which was dried
at 200\,$^{\circ}$C for 2\,h. The obtained product was crushed and transferred
to a corundum crucible. The first heating at 850\,$^{\circ}$C for 12\,h yielded
a grey-green product. Several annealing steps with grinding in between
followed. The final product was obtained after heat treatment at
1000\,$^{\circ}$C for 48\,h.

The powder X-ray diffraction pattern (Huber image plate Guinier camera G670, Ge
monochromator, Cu $K_{\alpha1}$ radiation, $\lambda$\,=\,1.5406\,\r{A},
powdered sample fixed with Vaseline between two Mylar foils each 6\,$\mu$m
thick) indicated the formation of single-phase \crbpo.\cite{crbpo_str_2000} 

Neutron powder diffraction data were collected\footnote{The $^{11}$B-containing
chemical (H$_3^{11}$BO$_3$) has been used to prepare a sample for neutron
powder diffraction.} using the BT-1 high-resolution powder diffractometer at
the NIST Center for Neutron Research. A monochromatic neutron beam with the
wavelength of 1.5403\,\AA\ was produced by a Cu~(311) monochromator.
Collimators with horizontal divergences of 15$^{\prime}$, 20$^{\prime}$, and
7$^{\prime}$ full width at half-maximum were used before and after the
monochromator, and after the sample, respectively. The intensities were
measured in steps of 0.05$^{\circ}$ in the 2$\theta$ range 3--168$^{\circ}$.
The data were collected at 4\,K, 35\,K, and 300\,K.  Additionally, the magnetic
scattering was studied with the triple-axis spectrometer BT-7 using the
wavelength of 2.359~\r A. The intensity of the strongest magnetic reflection
was monitored with the step of 0.5\,K in the $5-35$\,K temperature range.  The
structural analysis was performed using the program \texttt{GSAS}.\cite{GSAS}
The magnetic structure was refined with \texttt{Fullprof}.\cite{fullprof}

The magnetic susceptibility was measured using a commercial Quantum Design MPMS
SQUID in the temperature range 2--380\,K in magnetic fields up to 5\,T. The
electron spin resonance (ESR) measurement was performed at room temperature
with a standard continuous-wave spectrometer at X-band frequencies ($\nu\approx
9.5$~GHz) by using a cylindrical resonator in TE$_{012}$ mode.

DFT calculations have been performed using the full-potential local-orbital
code \textsc{fplo9.00-33} (Ref.~\onlinecite{FPLO}) and the pseudopotential
projector-augmented-wave code \textsc{vasp-5.2} (Ref.~\onlinecite{VASP,
*VASP_2}). For the scalar-relativistic calculations, we used the local density
approximation (LDA)\cite{PW92} and generalized gradient approximation
(GGA)\cite{PBE96} exchange-correlation potentials. Spin-unpolarized
calculations were performed on a 14$\times$14$\times$14 mesh of $k$-points. For
the spin-polarized calculations (DFT, DFT+$U$), we doubled the cell along $c$
and used a 4$\times$4$\times$2 $k$-mesh.  Hybrid-functional calculations were
performed in \textsc{vasp} using the HSE06 functional\cite{HSE03, *HSE04} on a
2$\times$2$\times$2 $k$-mesh.\footnote{Large computational demands incurred by
the hybrid functionals lead to severe restrictions on the number of $k$ points.
Nevertheless, the results obtained on a sparse $k$ mesh are sufficiently
accurate and well-converged, owing to the insulating nature of
\crbpo.} The convergence with respect to the $k$-meshes has been
accurately checked.  All calculations have been performed based on the 
crystal structure determined at 4\,K, as given in Table~\ref{T_str}. 

QMC simulations were performed using the code \textsc{looper} from the
software package \textsc{alps} version 1.3.\cite{ALPS} The magnetic
susceptibility was simulated on 8$\times$8$\times$64 finite lattices of
$S\!=\!3/2$ spins in the temperature range 0.25--8\,$J_1$, corresponding
to 12.5--400\,K (see Sec.~\ref{S-simul}) using 30\,000 sweeps for
thermalization and 300\,000 sweeps after thermalization. The statistical
errors ($<$0.5\%) are below the experimental accuracy. For simulations
of the spin stiffness and the static structure factor, we used finite
lattices up to 13\,824 and 2\,048 spins, respectively.

For the classical MC simulations, we used the \textsc{spinmc} code,\cite{ALPS}
with 200\,000 and 2\,000\,000 sweeps for and after thermalization,
respectively.  The length of the classical vectors is chosen such that the
maximal diagonal correlation matches the exact quantum result.  Chains of
$N\!=\!800$ spins were evaluated. Exact diagonalization of the $S^z\!=\!0$
sector was performed for $N\!=\!14$ sites alternating $S\!=\!3/2$ chain using
\textsc{sparsediag} from the \textsc{alps} package.\cite{ALPS}

\section{\label{S-exp}Experimental results}
\subsection{\label{S-cryst}Crystal structure}
The crystal structure of \crbpo\ has been initially refined from x-ray powder
data in the space group $P3$.\cite{crbpo_str_2000} However, the crystal structures of
related $M$(III)$_2[$BP$_3$O$_{12}]$ borophosphates were determined ($M$ =
In) or re-determined ($M$ = Fe) from single crystal data in the space group
$P6_3/m$.\cite{zhang2010,*li2010} This apparent mismatch
led us to reconsider the crystal structure of \crbpo. A thorough analysis of
both X-ray and neutron powder patterns identified the reflection condition
$00l$ with $l=2n$, which is characteristic of the $6_3$ screw axis.  No other
reflection conditions could be observed, so that the list of possible space
groups is restricted to $P6_3$, $P6_3/m$, and $P6_322$. While the refinement in
$P6_322$ was unsuccessful, the space group $P6_3/m$ resulted in low residuals and
a fully ordered crystal structure. Therefore, this centrosymmetric space
group was preferred over its non-centrosymmetric subgroup $P6_3$. No
significant structural changes between 4\,K and room temperature were detected.
Crystallographic data are listed in Table~\ref{T_str}. Selected interatomic
distances at 4\,K are given in Table~\ref{T_dist}.

\begin{figure}[tbp]
\includegraphics[width=8.6cm]{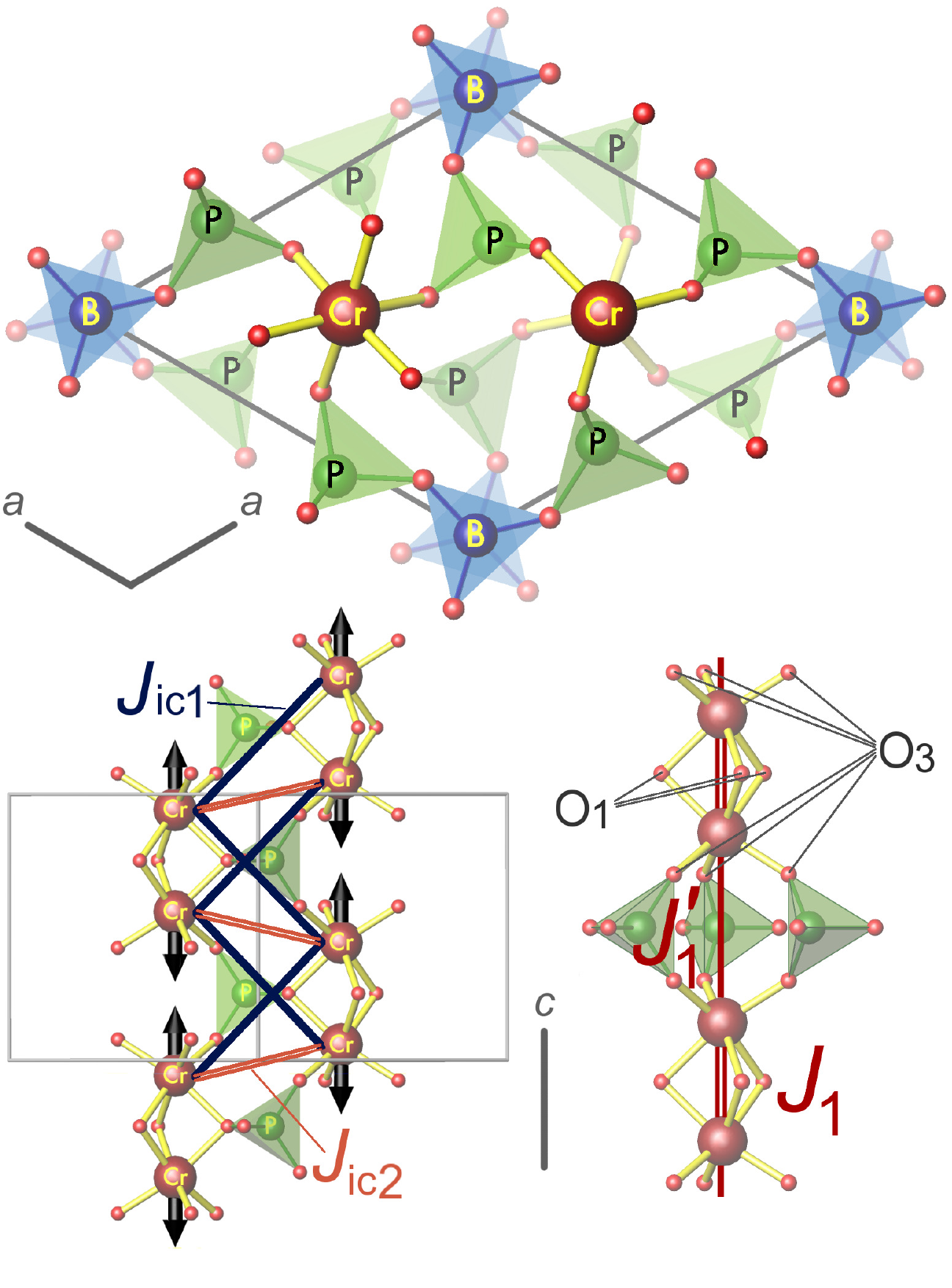}
\caption{\label{F-str}(Color online) Crystal structure of \crbpo. Top: 
view along $[0001]$.  Bottom left panel: view perpendicular to $[0001]$. 
The pathways of the leading interchain couplings
\jici\ and \jicii\ are shown with dark gray (dark blue) and light gray (red)
lines, respectively. Thick arrows denote the experimental magnetic structure refined 
in $\Gamma_1$ (see Sec.~\ref{S-magn} for details).
Bottom right panel: dimers are connected via three PO$_4$ tetrahedra and form
chains running along $[0001]$. The intra-dimer coupling $J_1$ (double
line) as well as the interdimer coupling $J_1'$ (single line) are shown.}
\end{figure}

The structure solution of \crbpo\ in the space group $P6_3/m$ is consistent with
earlier results for other $M_2[$BP$_3$O$_{12}]$ transition-metal
borophosphates.\cite{zhang2010,*li2010} Their crystal structures are isotypic and
feature dimers of face-sharing $M$O$_6$ octahedra (Fig.~\ref{F-str}, bottom). The
octahedra exhibit a sizable trigonal distortion leading to two groups of
nonequivalent Cr--O distances. The shorter Cr--O3 distances of about 1.91~\r A
(at 4\,K) take the terminal position of the Cr$_2$O$_9$ dimer, whereas the
longer Cr--O1 distances of about 2.04\,\r A are bridging (Fig.~\ref{F-str},
bottom right). The formation of longer Cr--O distances in the bridging position
is due to the Cr--Cr contact remaining relatively long (about 2.8\,\r A), in
order to reduce the repulsion between the positively charged Cr$^{3+}$ ions.
PO$_4$ tetrahedra link the Cr$_2$O$_9$ dimers along $[0001]$, whereas the
BO$_3$ triangles interconnect the PO$_4$ tetrahedra and do not share oxygen
atoms with the CrO$_6$ octahedra (Fig.~\ref{F-str}, top).

\begingroup
\squeezetable
\begin{table*}[tbp]
\caption{\label{T_str}Crystallographic data for \crbpo\ (space group $P6_3/m$)
according to the neutron powder diffraction data at 4, 35 and 300\,K.}
\begin{ruledtabular}
\begin{tabular}{c c | c c c | c c c | c c c}
\multicolumn{3}{l}{$T$}       & 4\,K & & \multicolumn{3}{c|}{35\,K} & \multicolumn{3}{c}{300\,K} \\
\multicolumn{3}{l}{$a$ (\AA)} & 7.9444(2) & & \multicolumn{3}{c|}{7.9448(2) } & \multicolumn{3}{c}{7.9524(2)} \\ 
\multicolumn{3}{l}{$c$ (\AA)} & 7.3439(3) & & \multicolumn{3}{c|}{7.3448(3)} & \multicolumn{3}{c}{7.3543(3)} \\ 
\multicolumn{3}{l}{$V$ (\AA$^3$)} & 401.40(2) & & \multicolumn{3}{c|}{401.49(3)} & \multicolumn{3}{c}{402.78(3) } \\ 
\multicolumn{3}{l}{$R_{\rm p}$ (\%)} & 5.5 & & \multicolumn{3}{c|}{5.8} & \multicolumn{3}{c}{6.4} \\
\multicolumn{3}{l}{$R_{\rm wp}$ (\%)} & 6.9 & & \multicolumn{3}{c|}{7.6} & \multicolumn{3}{c}{7.9} \\
\multicolumn{3}{l}{$R_{\rm exp}$ (\%)} & 6.1 & & \multicolumn{3}{c|}{7.5} & \multicolumn{3}{c}{8.4} \\\hline
& & $x/a$ & $y/b$ & $z/c$ & $x/a$ & $y/b$ & $z/c$ & $x/a$ & $y/b$ & $z/c$ \\
Cr & $4f$ & 1/3 & 2/3 & 0.0578(8) & 1/3 & 2/3 & 0.0614(9) & 1/3 & 2/3 & 0.0590(7) \\
P  & $6h$ & 0.3657(4) & 0.3166(5) & 1/4 & 0.3658(4) & 0.3168(5) & 1/4 & 0.3653(4) & 0.3159(5) & 1/4 \\
B  & $2a$ & 0 & 0 & 1/4 & 0 & 0 & 1/4 & 0 & 0 & 1/4 \\
O1 & $6h$ & 0.4023(4) & 0.5254(4) & 1/4 & 0.4009(4) & 0.5249(5) & 1/4 & 0.3997(4) & 0.5255(5) & 1/4\\
O2 & $6h$ & 0.1362(4) & 0.1931(4) & 1/4 & 0.1351(5) & 0.1924(4) & 1/4 & 0.1358(4) & 0.1926(4) & 1/4 \\
O3 & $12i$ & 0.4368(3) & 0.2691(3) & 0.0774(2) & 0.4365(3) & 0.2697(4) & 0.0768(3) & 0.4382(3) & 0.2716(3) & 0.0786(3) \\
\end{tabular}
\end{ruledtabular}
\end{table*}
\endgroup

\begin{table}
\caption{\label{T_dist}
Selected interatomic distances (in~\r A) in the \crbpo\ crystal structure at 4\,K. 
}
\begin{ruledtabular}
\begin{tabular}{cccc}
  Cr--O1 & $3\times 2.041(5)$ & P--O1 & 1.535(4)          \\ 
  Cr--O3 & $3\times 1.910(3)$ & P--O2 & 1.580(4)          \\
  Cr--Cr &   2.823(11)        & P--O3 & $2\times 1.511(2)$\\
  B--O2  & $3\times 1.365(3)$ &       &                   \\
\end{tabular}
\end{ruledtabular}
\end{table}
  
\subsection{\label{S-magn}Magnetic structure}
\begin{figure}[tbp]
\includegraphics[width=8.6cm]{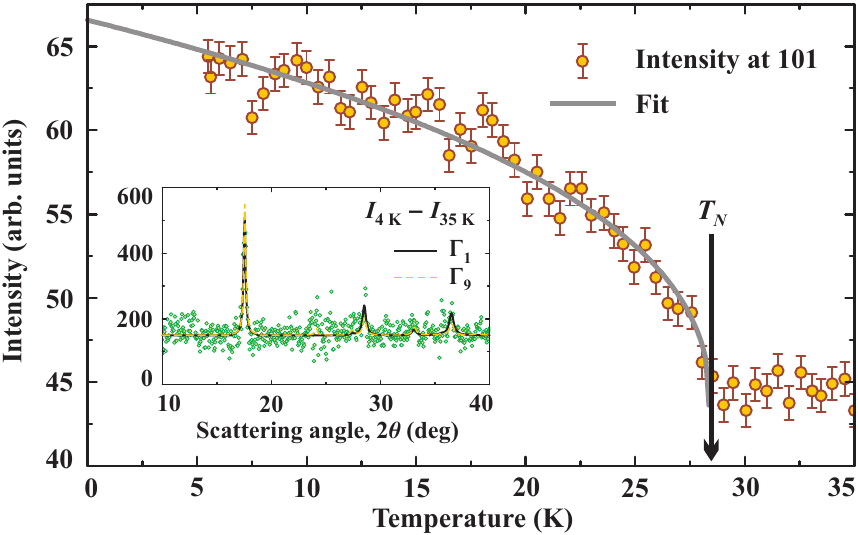}
\caption{\label{F_npd}(Color online) \crbpo: temperature evolution of the
magnetic 101 reflection (circles) and the fit with Eq.~\eqref{eq:neutron}. The
residual intensity above $T_N$ is due to the nuclear scattering and background.
Inset: refinement of the subtracted ($I_{\text{4\,K}}-I_{\text{35\,K}}$)
pattern with spins along $c$ ($\Gamma_1$, dark solid line) and along $a$ ($\Gamma_9$, 
light dashed line). Note that the temperature 
scan (main figure) and the angular scan (inset) are done on different instruments, 
hence the respective intensities should not be compared.  }
\end{figure}

At 4\,K, an additional magnetic scattering was observed. The only visible
magnetic reflection is at $2\theta\simeq 17.6\!^{\circ}$ and matches the weak 101
reflection of the nuclear structure. As the N\'eel temperature of \crbpo\ is
$T_N\simeq 28$\,K (see below) and no structural changes below $T_N$ are
expected, the subtraction of the 35\,K data from the 4\,K data results in the
purely magnetic scattering (Fig.~\ref{F_npd}, inset). However, no clear signatures of other
magnetic reflections could be observed. 

According to \texttt{Basireps}, the $P6_3/m$ space group, the $\kv=0$ propagation vector,
and the $4f$ Wyckoff position of Cr
allow for 12 irreducible representations with the magnetic moments lying in the $ab$ 
plane or pointing along the $c$ direction. Most of these representations can be
discarded because they produce the largest magnetic reflection at 100, 110, or 111,
which is contrary to the experimental observation of the magnetic scattering at 101
(Fig.~\ref{F_npd}, inset). The refinement is possible only in two representations, 
$\Gamma_1$ and $\Gamma_9$, that entail same ordering pattern (Fig.~\ref{F-str}, 
bottom left) with the magnetic moments along $c$ and $a$, respectively.

To evaluate the ordered magnetic moment, we refined the subtracted pattern as
the purely magnetic phase, and fixed the scale factor according to the
refinement of the 35\,K pattern, which is free from magnetic scattering. The
resulting magnetic moment at 4\,K is 2.5(1)~$\mu_B$ in $\Gamma_1$ and 1.1(1)~$\mu_B$
in $\Gamma_9$, with the refinement residuals of 0.131 in both models. While the refinements
within $\Gamma_1$ and $\Gamma_9$ are slightly different at high angles (see the 
inset of Fig.~\ref{F_npd}), the quality of the powder data is insufficient to 
observe these marginal differences and to discriminate between the two models. 
However, the spin-only magnetic moment of 3~$\mu_B$, which is expected for the $S=3/2$ Cr$^{3+}$ ion, 
strongly favors the solution in $\Gamma_1$. The somewhat lower experimental value of 2.5~$\mu_B$ 
is due to quantum fluctuations (see Sec.~\ref{S-simul}). 

The magnetic structure of \crbpo\ is shown in Fig.~\ref{F-str} (bottom left). We find that 
the Cr moments are antiferromagnetically ordered within each Cr$_2$O$_9$ dimer. The
interdimer ordering is also AF, both in the $ab$ plane and along the $c$ direction.

Temperature evolution of the magnetic moment can be tracked by the temperature 
dependence of the magnetic reflection 101 (Fig.~\ref{F_npd}). These data are fitted with 
the empirical formula 
\begin{equation}
I(T)=I_{\text{bg}}+I_0\left(1-\frac{T}{T_N}\right)^{\beta},
\label{eq:neutron}
\end{equation}
where $I_{\text{bg}}$ refers to the nuclear scattering and background above
$T_N$. The fit yields the N\'eel temperature $T_N=28.2(7)$\,K and the critical
exponent $\beta=0.38(3)$. The estimated $T_N\simeq 28$\,K is in excellent
agreement with the magnetic susceptibility measurement presented below
(Fig.~\ref{F_chi}), whereas $\beta$ falls into the range of values (0.36--0.39)
proposed for the 3D Heisenberg model.\cite{campostrini2002, *pelissetto2002}

\subsection{\label{S-chiT}Magnetic susceptibility and ESR} 
\begin{figure}[tbp]
\includegraphics[width=8.6cm]{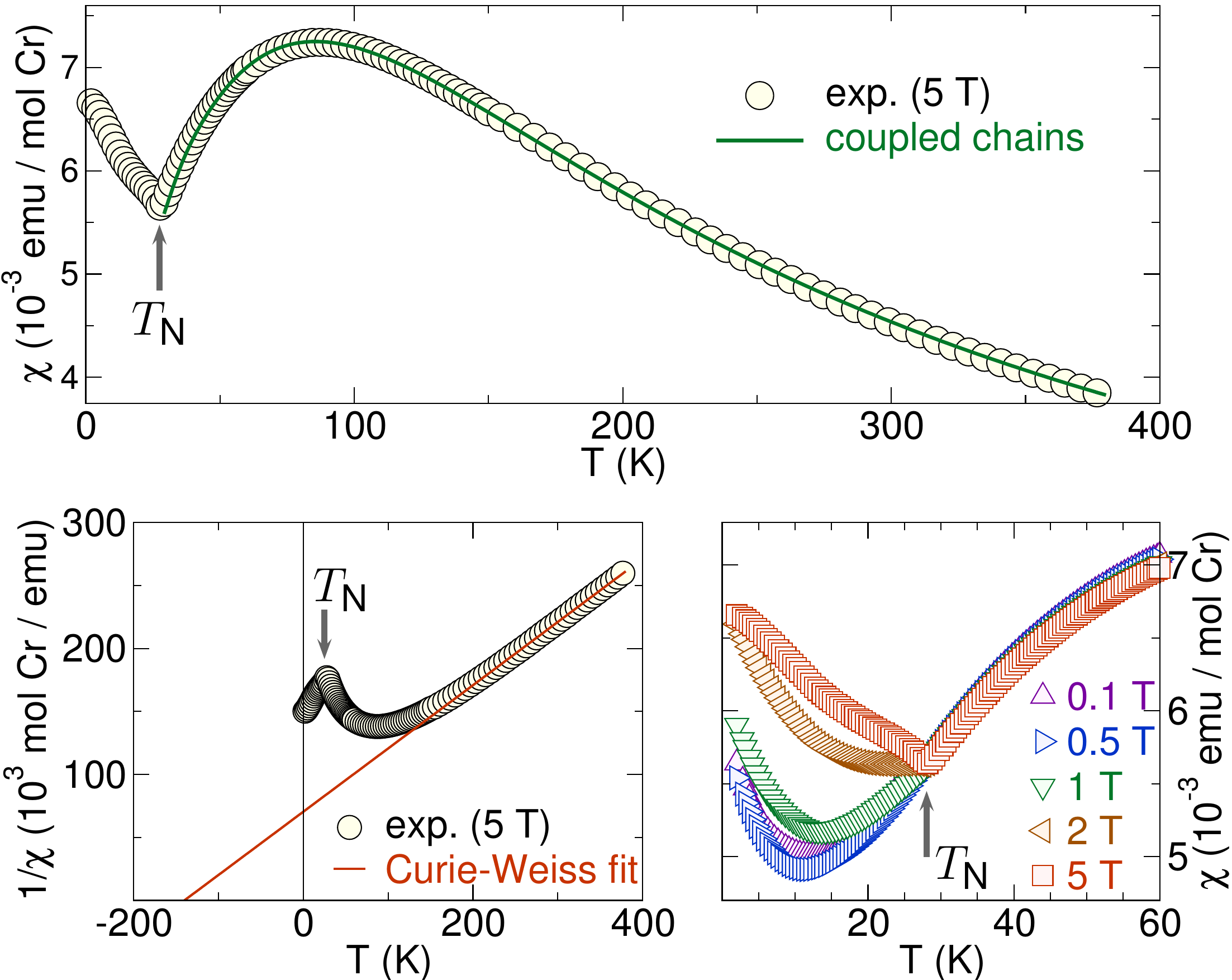}
\caption{\label{F_chi}(Color online) Top panel: Experimental
magnetic susceptibility $\chi(T)$ of \crbpo\ measured at 5\,T (circles) and fit
with the model of coupled bond-alternating $S$=3/2 chains (line).
Bottom left panel: Curie-Weiss fit to experimental $\chi(T)$. Bottom
right panel: Field dependence of $\chi(T)$. Note the long-range AF ordering
transition at $T_N$\,=\,28\,K.} 
\end{figure}

The temperature
dependence of the magnetic susceptibility $\chi(T)$ (Fig.~\ref{F_chi}, top)
reveals a typical low-dimensional behavior with a broad maximum around
$T$\,=\,85\,K. The high-temperature part ($T$\,$>$\,200\,K) of the curve obeys
the Curie-Weiss law (Fig.~\ref{F_chi}, bottom left) with the Curie constant
$C$\,=\,1.987\,emu\,K\,(mol Cr)$^{-1}$ leading to the effective magnetic moment
$\mu_{\rm
eff}$\,=\,$\sqrt{3\,C\,k_{\text{B}}\,\mu_{\text{B}}^{-2}\,N_{\text{A}}^{-1}}$\,=\,3.987\,$\mu_B$
per Cr, which is slightly larger than the spin-only contribution $\mu_{\rm
eff}$\,=\,$g\sqrt{S(S+1)}$\,$\simeq$\,3.88\,$\mu_{\text{B}}$ (assuming the
orbital moment is completely quenched). 

The Weiss temperature $\theta$\,=\,139\,K and the broad maximum
around $T^{\rm max}$\,=\,85\,K, corresponding to $\sim$61\% of $\theta$,
indicate sizable AF spin correlations. The magnetic ordering
temperature $T_{N}$\,=\,28\,K can be clearly traced by the divergence of the
$\chi(T)$ curves measured in low and high fields. While the low-field
measurements show only an inconspicuous bend around $T_N$, the measurements
above 2~T reveal a well-defined cusp. This divergence is due to the spin-flop
transition in the AF-ordered phase. The spin-flop transition indeed takes place
at about 1.7~T, as shown by the low-temperature magnetization
curve.\cite{supplement}

Room-temperature ESR measurements yield a narrow line that can be fitted with a
single powder-averaged Lorentzian.\cite{supplement} Within resolution, the
resonance field appears isotropic and corresponds to the $g$ value of 1.968
which indicates weak spin-orbit coupling, as typical for Cr$^{3+}$.

The low-dimensional magnetism of \crbpo\ is consistent with the presence of
Cr$_2$O$_9$ dimers in the crystal structure. However, isolated spin dimers
should have a singlet GS at low temperatures. The fact that \crbpo\
develops long-range AF order with a sizable N\'eel temperature of 28\,K implies
substantial interdimer couplings via the PO$_4$ groups. While such couplings
are abundant in vanadium and copper
phosphates,\cite{garrett1997,tsirlin2008,janson2011,*lebernegg2011} their
identification is by no means a simple task. For example, in \crbpo\ the
shortest interdimer Cr--Cr distances are 4.55~\r A along $[0001]$ and
4.67~\r A in the $(0001)$ plane, and it is impossible to decide \textit{a priori}
whether one of these pathways is more efficient, or both should be treated on
equal footing. To develop a reliable magnetic model of \crbpo, we perform
extensive DFT calculations followed by QMC simulations, and provide detailed
microscopic insight into the magnetism of this compound.

\section{\label{S-DFT}DFT calculations}
To unravel the microscopic origin of the observed low-dimensional magnetic
behavior, we apply DFT calculations and evaluate the individual exchange
couplings.  Nonmagnetic (spin-unpolarized) calculations within both LDA and GGA
yield a well-structured valence band (Fig.~\ref{F_dos_bs}, top) with the total
width of $\sim$10\,eV, crossed by the Fermi level $\varepsilon_F$ at zero
energy. This metallic electronic GS, contrasting with the green color of
\crbpo, originates from the underestimation of the strong electron-electron
repulsion within the Cr $3d$ shell.  Technically, the band gap can be readily
restored in a spin-polarized calculation, due to the sizable exchange splitting
typical for Cr$^{3+}$. However, its value is too small (1.29\,eV in LSDA,
1.78\,eV in GGA)\cite{supplement} to account for the green color
($\simeq$2.2--2.5\,eV) of \crbpo.  Accounting for electronic correlations is
challenging, since many-body effects can not be properly described within the
one-electron approach of conventional DFT functionals.

Typically, multiorbital correlated insulators are described by an extended
Hubbard model that comprises the kinetic terms (electron transfer), the on-site
and intersite correlations (Coulomb repulsion), and the on-site exchange
(Hund's exchange). However, the large Hilbert space of such models impedes even
an approximate numerical solution that would establish a simple relation
between the parameters of the Hubbard model and the resulting magnetic
couplings. Therefore, we restrict ourselves to a qualitative analysis of
electron transfers. 

For the quantitative evaluation of magnetic couplings, we account for
correlation effects using the mean-field-like DFT+$U$ approach or the hybrid (DFT +
exact exchange) functionals. The comparison to the experiment demonstrates the
good agreement between different computational approaches, and underscores the
validity of the qualitative analysis based on the electron transfers. Thus,
this qualitative model approach could also be used as a starting point for more
involved Cr$^{3+}$ systems with nontrivial coupling pathways.

\subsection{Qualitative model approach}
\label{S_model}
Conventional DFT (LDA and GGA) functionals
are known for their accurate description of electron transfer processes,
thus kinetic terms of the model Hamiltonian can be evaluated directly
from the LDA/GGA band dispersions for the Cr $3d$ states.  Typical for
an octahedral environment, these states split into two manifolds:
half-filled $t_{2g}$ orbitals centered at $\varepsilon_F$, and empty
$e_g$ orbitals that lie $\sim$1.7\,eV higher in energy owing to the
sizable crystal-field splitting. 

\begin{figure}[tbp]
\includegraphics[width=8.6cm]{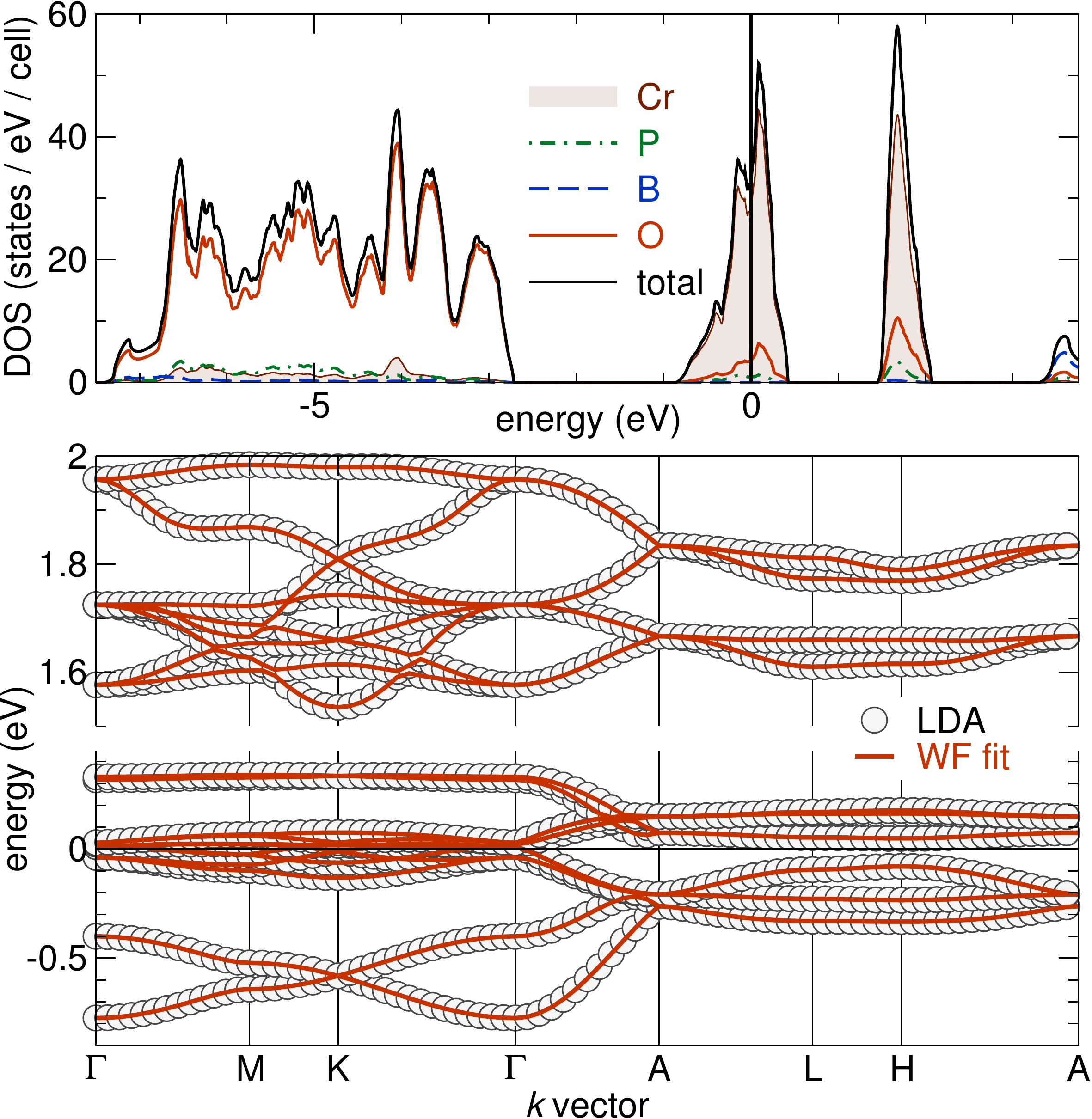}
\caption{\label{F_dos_bs}(Color online) Electronic structure of \crbpo. Top:
LDA density of states. The Fermi level $\varepsilon_F$ is at zero
energy. The contribution from Cr states is shown by shaded filling. Bottom: LDA
bands in the vicinity of the Fermi level and the fit with the tight-binding
model based on Wannier functions.}
\end{figure}

\begin{figure*}[tb]
\includegraphics[width=17.8cm]{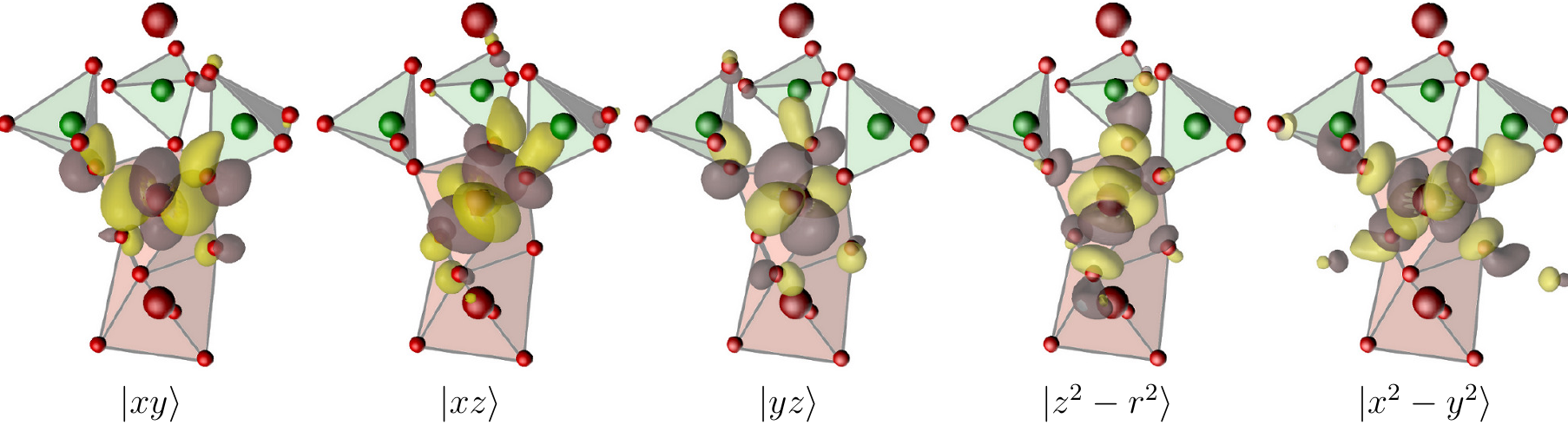}
\caption{\label{F_wf}(Color online) \crbpo: Wannier functions for five Cr $3d$
orbitals (see subscripts). Cr, P and O atoms are depicted by large (red),
intermediate-size (green) and small (red) spheres, respectively. For each of
the five Wannier functions, a Cr$_2$O$_9$ dimer and three PO$_4$ tetrahedra are
shown. The projection is similar to the bottom right panel of
Fig.~\ref{F-str}.}
\end{figure*}

Altogether, the $t_{2g}$ and $e_g$ manifolds comprise 20 bands, in line with
five $3d$ orbitals per Cr atom and four Cr atoms in a unit cell. To evaluate
the transfer integrals $t^{mm'}_{ij}$ ($m$ and $m'$ are orbital indices, $i$
and $j$ are site indices), we map these 20 bands onto an effective five-orbital
tight-binding model, and parameterize this model using the Wannier functions
(WF) technique.\footnote{Here, we do not adapt the local coordinate frame to
the global symmetry of the crystal structure, and direct the local $z$ axis
along one of the Cr--O bonds (see also Fig.~\ref{F_wf}). Therefore, the
resulting $d$ states do not show the weak splitting in the $t_{2g}$ subspace,
as expected for trigonally distorted octahedron. This simplification does not
affect any of our results, because Cr$^{3+}$ shows the robust high-spin state
with the half-filled $t_{2g}$ levels, whereas fine structure of these levels
has minor effect on the magnetism.} In the WF basis (Fig.~\ref{F_wf}), the
couplings $t^{mm'}_{ij}$ are evaluated as nondiagonal matrix elements. The
resulting $t^{mm'}_{ij}$ show excellent agreement between the tight-binding
model and the computed LDA/GGA band dispersions (Fig.~\ref{F_dos_bs}, bottom).

\begin{table}[tbp]
\caption{\label{T_WF} Leading transfer integrals $t_i^{mm'}$ (notation
according to Fig.~\ref{F-str}), where $m$ and $m'$ are orbital indices from the
following set: $|xy\rangle$, $|xz\rangle$, $|yz\rangle$, $|z^2-r^2\rangle$, and
$|x^2-y^2\rangle$. All values are given in~meV. For clarity, only one of two
symmetrically equivalent terms $t_{ij}^{mm'}$ and $t_{ij}^{m'm}$ is shown. The
Cr--Cr distances ($d_{\text{Cr--Cr}}$) are given for the 4\,K structure
(Table~\ref{T_str}).}
\begin{ruledtabular}
\begin{tabular}{r r r r r r}
\multicolumn{6}{c}{$t_1$}\\
\multicolumn{6}{c}{($d_{\text{Cr--Cr}}$\,=\,2.823\,\r{A})}\\
& $|xy\rangle$ & $|xz\rangle$ & $|yz\rangle$ & $|z^2-r^2\rangle$ & $|x^2-y^2\rangle$\\ 
$\langle{}xy|$      & $-38$  & $-125$ & $-125$& $-97$  & -- \\ 
$\langle{}xz|$      &        & $-38$  & $-125$& 39     & $-89$\\
$\langle{}yz|$      &        &        & $-38$ & 58     & 78.9 \\
$\langle{}z^2-r^2|$ &        &        &       & --     & -- \\
$\langle{}x^2-y^2|$ &        &        &       &        & --\\ \hline
\multicolumn{6}{c}{$t_1'$}\\
\multicolumn{6}{c}{($d_{\text{Cr--Cr}}$\,=\,4.521\,\r{A})} \\
& $|{}xy\rangle$ & $|{}xz\rangle$ & $|yz\rangle$ & $|{}z^2-r^2\rangle$ & 
$|{}x^2-y^2\rangle$ \\
$\langle{}xy|$      & $-124$ & --     & --    & --     & -- \\
$\langle{}xz|$      &        & $-124$ & --    & 33     & -- \\
$\langle{}yz|$      &        &        & $-124$& --     & -- \\
$\langle{}z^2-r^2|$ &        &        &       & $-60$  & -- \\
$\langle{}x^2-y^2|$ &        &        &       &        & $-60$\\ \hline
\multicolumn{6}{c}{$t_{\rm ic1}$}\\
\multicolumn{6}{c}{($d_{\text{Cr--Cr}}$\,=\,5.875\,\r{A})} \\
& $|{}xy\rangle$ & $|{}xz\rangle$ & $|yz\rangle$ & $|{}z^2-r^2\rangle$ & 
$|{}x^2-y^2\rangle$ \\
$\langle{}xy|$      & --     & --     & --    & 86    & $-60$\\
$\langle{}xz|$      &        & --     & --    & $-44$ & 75 \\
$\langle{}yz|$      &        &        & --    & $-67$ & --\\
$\langle{}z^2-r^2|$ &        &        &       & 30    & 35 \\
$\langle{}x^2-y^2|$ &        &        &       &       & 38 \\ \hline
\multicolumn{6}{c}{$t_{\rm ic2}$}\\ 
\multicolumn{6}{c}{($d_{\text{Cr--Cr}}$\,=\,4.665\,\r{A})} \\
& $|{}xy\rangle$ & $|{}xz\rangle$ & $|yz\rangle$ & $|{}z^2-r^2\rangle$ & 
$|{}x^2-y^2\rangle$ \\
$\langle{}xy|$      & --     & $-30$  & $-30$ & --    & --\\
$\langle{}xz|$      &        & --     & --    & --    & --\\
$\langle{}yz|$      &        &        & --    & --    & --\\
$\langle{}z^2-r^2|$ &        &        &       & --    & --\\
$\langle{}x^2-y^2|$ &        &        &       &       & $-31$\\ 
\end{tabular}
\end{ruledtabular}
\end{table}

Sizable transfer integrals are found for four Cr--Cr pathways
(Table~\ref{T_WF}): the intradimer transfer ($t_1$), the interdimer tranfer
along $c$ ($t_1'$), the shortest interdimer pathway in the $ab$ plane
($t_{ic2}$), and the longer interdimer pathway along the $[2\bar 23]$ direction
($t_{ic1}$), see Fig.~\ref{F-str}. Here, $t_1'$ and $t_{ic2}$ run via triple
bridges of the PO$_4$ polyhedra (along $c$), whereas $t_{ic1}$ runs via a
single PO$_4$ tetrahedron. Other couplings are negligible, as they involve the
transfer through at least two PO$_4$ tetrahedra in a row.

Although an explicit expression relating exchange integrals $J_{ij}$ to
transfer integrals $t_{ij}^{mm'}$ is presently not available, the qualitative
comparison of $t_{ij}^{mm'}$ terms elucidates different contributions to the
magnetic exchange. According to the Goodenough-Kanamori rules, the hoppings between
the half-filled states (both $m$ and $m'$ belong to the $t_{2g}$ subspace) are
responsible for AF couplings, whereas the hoppings between the half-filled and
empty states ($m$ belongs to $t_{2g}$, $m'$ belong to $e_g$) give rise to
ferromagnetic (FM) interactions. 

First, we consider the intradimer $t^{mm'}_1$ hoppings, where the couplings
between different $t_{2g}$ orbitals are dominant:
$t^{xy,xz}_{1}\!=\!t^{xz,yz}_{1}$\,=\,$t^{yz,xy}_{1}\!=\!-125$\,meV. At the
same time, the largest hopping between the $t_{2g}$ and $e_g$ orbitals is
smaller ($t^{xy,z^2-r^2}_{1}\!=\!-97$\,meV), thus hinting at the AF nature of
$J_1$.  For the interdimer coupling $t_1'$ the difference is even more
pronounced: the hoppings between $t_{2g}$ orbitals amount to $-124$\,meV, thus
dominating over the hopping to the empty $z^2-r^2$ orbital.  Therefore, the
interdimer exchange $J_1'$ should be also AF.

The interchain coupling $J_{\rm ic2}$ is realized primarily via $t^{xy,xz}_{\rm
ic2}$ and $t^{xy,yz}_{\rm ic2}$ hoppings, hence the AF contribution is again
dominant.  In contrast, for the interchain coupling $J_{\rm ic1}$, the
$t_{2g}\rightarrow t_{2g}$ hoppings are close to zero, whereas the
$t_{2g}\rightarrow e_g$ hoppings are still sizable. Therefore, the overall
exchange should be FM.

The above qualitative analysis is confirmed by elaborate DFT+$U$ and
hybrid-functional calculations reported below. More importantly, the proposed
couplings are in agreement with the experimental magnetic structure that
features antiparallel spins along $J_1$, $J_1'$, and $J_{\rm ic2}$
(Fig.~\ref{F-str}, bottom left panel). The respective AFM couplings establish
parallel spins along the $J_{\rm ic1}$ bonds that are indeed FM. Therefore, the
spin lattice of \crbpo\ is non-frustrated. Because $J_1'\gg J_{\rm ic1},J_{\rm ic2}$,
this spin lattice can be considered as quasi-1D, with bond-alternating
$J_1-J_1'$ chains running along the $c$ direction.

\subsection{Total-energy calculations} 
A sizable exchange splitting typical for Cr$^{3+}$ compounds readily opens a
band gap (albeit underestimating the experimental value, see
Ref.~\onlinecite{supplement}) in the spin-polarized calculations, so that the
magnetic energy can be evaluated directly from LSDA or GGA total-energy
calculations. In particular, total energies $E_{\rm tot}$ corresponding to
different collinear magnetic configurations can be mapped onto classical
Heisenberg model, thus yielding the exchange integrals $J_{ij}$:

\begin{equation}
\label{E:classHeis}E_{\rm tot}=E_0+E_{\rm magn}=E_0+\sum_{\langle{}i,j\rangle}J_{ij}\vec{S}_i\cdot\vec{S}_j.
\end{equation}

We apply this approach to evaluate the intrachain and interchain couplings. In
LSDA, we obtain $J_1\!=\!151$\,K and $J_1'\!=\!82$\,K as well as the FM
\jici\,=\,$-8$\,K and the AF \jicii\,=\,2\,K. Spin-polarized GGA yields
marginally smaller values (Table~\ref{T-J}). To challenge these estimates, we
address the Weiss temperature $\theta$, which is a linear combination of
magnetic couplings.\cite{HC_AHC_Johnston} For the \crbpo\ spin lattice, the
expected Weiss temperature is:
\begin{equation}
\label{E:CW}\theta=\frac{1}{3}S(S+1)(J_1+J_1'+6J_{\rm ic1}+3J_{\rm ic2}),
\end{equation}
where each coupling is summed up according to its coordination number in the
spin lattice. Adopting the values of $J_i$ from LSDA (GGA), we obtain
$\theta\!\simeq$\,242\,K (202\,K), which is 50\,--\,70\% larger than the
experimental value $\theta\!=\!139$\,K. 

This discrepancy primarily originates from the poor description of
electronic correlations within LDA or GGA, and calls for the application of more elaborate
computational approaches. For strongly correlated insulators, such as
\crbpo, the mean-field DFT+$U$ method is a natural choice.  An intrinsic problem
of this method is the double counting (DC) of the correlation energy
already present in LDA or GGA. This correlation energy should be subtracted from the
total energy of the system. In the widely used DC corrections, denoted
around-mean-field (AMF) and fully-localized limit (FLL),\cite{LDA_U_AMF_FLL}
the subtracted energy corresponds to the energy of the uniformly occupied state
or the state with integer occupation numbers, respectively.  Previous DFT+$U$
studies on Cr$^{3+}$ materials\cite{LiCrO2_DFT_2007} did not render any of the two
schemes preferable, thus we apply both and compare their results with the
experiment.

Besides the DC, the DFT+$U$ method introduces two free parameters: the
on-site repulsion $U_d$ and the on-site exchange $J_d$. While an
empirical evidence favors $J_d\!\simeq\!1$\,eV for $3d$ elements, the
values of $U_d$ can substantially vary depending on the electronic
configuration and local environment of the magnetic ion. For Cr$^{3+}$,
we varied $U_d$ in the range between 2 and 4\,eV.\cite{LiCrO2_DFT_2007,fennie2006}

The resulting DFT+$U$-based exchange integrals are listed in Table~\ref{T-J}.
We find that the exchange integrals are weakly dependent on the DFT
exchange-correlation potential (LSDA or GGA). In contrast, the DC correction
plays a more substantial role: the calculations within AMF yield considerably
larger couplings than the calculations within FLL for the same value of $U_d$.
Besides, \jici\ turns out to be AF in AMF, but FM in FLL. This conspicuous
difference between different flavors of DFT+$U$ necessitates an additional
examination of this problem by an independent technique.

To this end, we resort to the HSE06 hybrid functional that is free from the DC
problem. Here, the DFT exchange is mixed with a fraction of exact
(Hartree-Fock) exchange.\cite{franchini2007,*iori2012} The HSE06-based exchange
integrals (Table~\ref{T-J}) are very similar to those obtained using DFT+$U$
FLL with $U_d\!=\!3.0$\,eV. To check the stability of the HSE06 results, we
varied the parameter $\beta$, which reflects the admixture of the exact
exchange to the standard DFT exchange ($\beta\!=\!0.25$ in the original
definition of HSE06). As follows from the resulting values (Table~\ref{T-J}),
the increase in $\beta$ from 0.2 to 0.3 is accompanied by the $\sim$25\%
reduction in the exchange couplings, while the $J_1'/J_1$ ratio is essentially
unchanged. Therefore, the effects of increasing $\beta$ in HSE06 and increasing
$U_d$ in DFT+$U$ are somewhat similar, despite the disparate physical meaning
of the $U_d$ and $\beta$ parameters.

On a qualitative level, DFT+$U$ FLL, hybrid functionals, and model
approach concur with each other on the nature of magnetic couplings in \crbpo.
By contrast, DFT+$U$ AMF predicts the different sign of $J_{\rm ic1}$, which would
render the spin lattice weakly frustrated. However, this scenario looks very
unlikely because the independent model analysis (Sec.~\ref{S_model}) yields FM
$J_{\rm ic1}$. Additionally, DFT+$U$ AMF systematically overestimates the absolute
values of $J_1$ and $J_1'$ (see experimental values in Sec.~\ref{S-simul}) and
should probably be discouraged in the case of \crbpo.\cite{[{It is widely
believed that for insulators only the FLL DC correction is appropriate. By the
same token, thorough studies of magnetic couplings in Cu and V oxides show that
AMF may provide comparable or even better results than FLL, although in a
rather unsystematic fashion: see, e.g., }][{}]tsirlin2010,*tsirlin2011}

Apart from the overall energy scale of $J_1$ and $J_1'$ that is dependent on
$U_d$ or $\beta$ (Table~\ref{T-J}), different flavors of DFT+$U$ as well as
HSE06 predict different $J_1'/J_1$ ratios. Our computational results yield
$J_1'/J_1=0.39\!-\!0.94$, so that $J_1'$ is smaller than $J_1$ but certainly
large enough to ensure the quasi-1D nature of the spin lattice. Unfortunately,
the more precise evaluation of the $J_1'/J_1$ ratio lies beyond the
capabilities of present-day DFT-based techniques and should be addressed by
numerical simulations of the spin model and subsequent fitting to the
experimental data (see Sec.~\ref{S-simul}).

\begin{table}[tbp]
\caption{\label{T-J}Leading exchange integrals $J_i$ (in K)
calculated with different DFT-based methods. For the DFT+$U$ results, the
double-counting correction (DCC) and $U_d$ value (in~eV) are given. For the
HSE06 hybrid functional, the admixture of the exact exchange ($\beta$) is
specified.}
\begin{ruledtabular}
\begin{tabular}{c c r r r r r}
$E_{\rm xc}$ (DCC) & $U_d$ or $\beta$ & $J_1$ & $J_1'$ & \jicii & \jici & $J_1'/J_1$
\\ \hline
LSDA  & &               150 & 82 & 2.5 & $-7.8$ & 0.55 \\
GGA & &                 125 & 63 & 2.2 & $-5.6$ & 0.51 \\ \hline 
 
              & 2.0 &   109 & 60 & 2.3 &  0.1   & 0.62 \\
LSDA+$U$ (AMF)& 3.0 &    89 & 55 & 2.2 &  1.7   & 0.72 \\
              & 4.0 &    70 & 50 & 2.1 &  2.8   & 0.86 \\ \hline
  
              & 2.0 &    84 & 49 & 1.6 & $-2.1$ & 0.58 \\
LSDA+$U$ (FLL)& 3.0 &    54 & 38 & 1.2 & $-1.8$ & 0.71 \\
              & 4.0 &    33 & 31 & 0.9 & $-1.6$ & 0.94 \\ \hline

              & 2.0 &    97 & 48 & 2.2 & 1.0    & 0.50 \\
GGA+$U$ (AMF) & 3.0 &    81 & 44 & 2.1 & 2.3    & 0.54 \\
              & 4.0 &    67 & 41 & 2.1 & 3.3    & 0.61 \\ \hline

              & 2.0 &    78 & 40 & 1.6 & $-1.0$ & 0.51 \\ 
GGA+$U$ (FLL) & 3.0 &    54 & 32 & 1.2 & $-1.0$ & 0.59 \\ 
              & 4.0 &    37 & 26 & 1.0 & $-0.9$ & 0.69 \\ \hline
        
             & 0.20 &    66 & 27 & 1.2 & $-2.5$ & 0.41 \\
HSE06        & 0.25 &    58 & 23 & 1.0 & $-2.0$ & 0.40 \\
             & 0.30 &    51 & 20 & 0.9 & $-1.7$ & 0.39
\end{tabular}
\end{ruledtabular}
\end{table}

Finally, we take into account the spin-orbit (SO) coupling, and quantify the
single-ion anisotropy $D$ using the DFT+$U$+SO calculations within the FLL in
\textsc{vasp}. To this end, we calculate total energies
(Fig.~\ref{F-str}, bottom left) of the AF GS with the magnetic moments aligned
parallel and perpendicular to the hexagonal axis ($c$-axis). For
$U_d\!=\!3$\,eV, the energy difference of 1.3\,K slightly favors the in-plane spin
arrangement. This result implies a very weak single-ion anisotropy. However, the
direction of the magnetic moment should be determined experimentally because
the energy difference of 1\,K is on the verge of the accuracy of DFT.

In summary, our band structure calculations arrive at a scenario of Heisenberg
spin chains with two alternating nearest-neighbor AF interactions $J_1$ and
$J_1'$. The magnetic chains are weakly coupled in a non-frustrated manner by
the FM \jici\ and the AF \jicii. While both interchain couplings are apparently
weak (below 2\,K in terms of the absolute value), the energy scale of the
intrachain couplings, as well as the precise alternation ratio, are rather
sensitive to details of the computational procedure and require further
refinement applying simulation techniques to fit the experimental data.

\section{\label{S-simul}QMC simulations and comparison with the
experiments}

To refine the parameters of the microscopic magnetic model, we simulate the
temperature dependence of the reduced magnetic susceptibility $\chi^{*}$ for
various ratios of $J_1'/J_1$, while keeping the interchain couplings constant
$\jicii\!=\!-\jici\!=\!0.02\!J_1$ (see Table~\ref{T-J}), and subsequently fit
the simulated QMC curves to the experiment using the expression:

\begin{equation}
\chi(T)=
\frac{N_Ag^2{\mu}_{B}^2}{k_B\,J_1}\cdot\chi^{*}\biggl(\frac{T}{k_B\,J_1}\biggl)
+ \chi_0.
\end{equation}

Here, the fitted parameters are $J_1$, the Land\'e factor $g$, and the
temperature-independent contribution $\chi_0$.\footnote{For the fitting, we
used the experimental data above $T_N$\,=\,28\,K, since a one-dimensional model
cannot account for the AF ordering
(Ref.~\onlinecite{Mermin_Wagner}).}\nocite{Mermin_Wagner} The results are
presented in Table~\ref{T-QMC} (columns 2 and~3). A general trend is the
decrease in $J_1$ upon the increase in $J_1'/J_1$, so that the sum of the
leading couplings $J_1+J_1'$ remains nearly constant ($74.5\pm2.0$\,K), see
also Eq.~\eqref{E:CW}. Still, smaller values of $J_1'/J_1$ yield better
agreement with the experimental curve, especially around the maximum in
$\chi(T)$.

\begin{table}[tbp]
\caption{\label{T-QMC}Results of QMC simulations and fitting to the
experimental magnetic susceptibility. $J_1$ and $g$ are obtained from the fits to $\chi(T)$. $T_{\rm N}/J_1$
is obtained from simulations of spin stiffness (see text). $S_{\infty}$
is evaluated using Eq.~(\ref{E:mSqN-scaling}). $T_{\rm N}$ and
$m$ are scaled using $J_1$ and $g$ from columns 2 and 3.}
\begin{ruledtabular}
\begin{tabular}{c c c c c c c}
$J_1'/J_1$ & $J_1$ (K) & $g$ & $T_{\rm N}/J_1$ & $T_{\rm N}$ (K) &
$S_{\infty}$ & $m (\mu_{\rm B})$ \\ \hline
0.4 & 51.84 & 2.006 & 0.584 & 30.27 & 1.128 & 2.22 \\
0.5 & 49.51 & 2.002 & 0.642 & 31.79 & 1.151 & 2.27 \\
0.6 & 47.10 & 1.996 & 0.689 & 32.45 & 1.164 & 2.29 \\
0.7 & 44.78 & 1.993 & 0.726 & 32.51 & 1.169 & 2.30 \\
0.8 & 42.58 & 1.992 & 0.755 & 32.15 & 1.165 & 2.29 \\
\end{tabular}
\end{ruledtabular}
\end{table}

In general, additional restrictions for the model parameters can be set by
the fitted values of $g$.  However, $g$ is marginally dependent on
$J_1'/J_1$ and shows good agreement with the ESR value of 1.968 in the whole
range studied.  Therefore, simulation of further measurable quantities
is needed to refine the model parameters. 

First, we address the magnetic ordering temperature. To this end, we calculate
the spin stiffness $\rho$ as a function of temperature for finite lattices with
up to $N\!=\!L_x\times{}L_y\times{}L_z$\,=\,13\,824 spins. At the magnetic
ordering transition temperature, the products $\rho_iL_i (i\!=\!x,y,z)$ are
independent of $L_i$. Thus, $T_{\rm N}$ is the temperature at which the
$\rho{}L(T)$ curves of the different finite lattices cross. The difference
between the simulated values of $T_{\rm N}$ and its experimental value of 28\,K
is below 10\% for the whole range of $J_1'/J_1$ (see columns 4 and 5 of
Table~\ref{T-QMC}). Similar to the fits of $\chi(T)$, lower values of
$J_1'/J_1$ yield slightly better agreement with the experimental $T_{\rm
N}\!=\!28$\,K. 

Another measured quantity, which can be used for comparisons between theory and
experiment, is the ordered magnetic moment $m$.  In the classical $S\!=\!3/2$
Heisenberg model, the ordered magnetic moment $m\!=\!g\mu_{\rm B}S$ amounts to
3\,$\mu_{\rm B}$, but the experimentally observed moment in \crbpo\ is
substantially smaller (2.5\,$\mu_{\rm B}$, see Sec.~\ref{S-magn}).  We estimate
the reduction in $m$ due to quantum fluctuations. To this end, we simulate the
magnetic structure factors $\mathbb{S}$ for finite lattices containing up to
2048 spins. 

For the propagation vector
$\kv$ of the magnetically ordered GS, the ordered magnetic moment
$m$ is evaluated using the finite-size scaling procedure:
\begin{subequations}
\label{E:mSqN-scaling}
\begin{eqnarray}
\frac{3{\mathbb S}(\kv)}{N}\,=\,S_{\infty}(\kv)^2
+\frac{\sigma_1}{\sqrt{N}} + \frac{\sigma_2}{N}\\
m=g\mu_{\rm B}S_{\infty}(\kv),
\end{eqnarray}
\end{subequations}
where the fitting parameters are $S_{\infty}(\kv)$ as well as the expansion
coefficients $\sigma_1$ and $\sigma_2$, while $g$\,=\,1.968 is adopted from the
ESR.\footnote{Unfortunately, the quasi-1D character of the magnetic model
requires $L_z\gg{}L_x\simeq{}L_y$ for an $N$-site lattice
($N\!=\!L_x\times{}L_y\times{}L_z$). This leads to a small number of
computationally feasible lattices, impeding an accurate estimation of the
magnetic moment. To check the results for consistency, we repeated the fitting
using a simplified scaling $\sigma_2\!=\!0$. As expected, the $\sigma_1$ values
are substantially renormalized. Yet, both approaches yield marginally different
values of $S_{\infty}$ ($\sim$1\,\% difference).} In this way, we arrive at the
values that underestimate the experimental result by $\simeq0.2$\,$\mu_{\rm B}$
(last column of Table~\ref{T-QMC}).  Since this offset is nearly independent of
$J_1'/J_1$, the interchain couplings might be the origin of this discrepancy.
Indeed, increasing the interchain couplings by a factor of two
(\jicii\,=\,$-\jici$\,=\,0.04\,$J_1$) yields $m$\,=\,2.49\,$\mu_{\rm B}$, in
excellent agreement with the experiment.  However, larger interchain couplings
substantially increase the N\'eel temperature and worsen the agreement between
the simulated and experimental $\chi(T)$. Since the experimental value of
$T_{\rm N}$ is more precise than $m$, we argue that the parameters
$\jicii\!=\!-\jici\!=\!0.02$\,$J_1$ are preferable, while the 0.2\,$\mu_{\rm
B}$ deviation between the experimental and simulated values of $m$ is still
reasonable. Indeed, the experimental estimate of $m$ relies upon the available
magnetic form-factors as well as on the accuracy of the subtraction procedure.
Therefore, we can not exclude a systematic experimental error in $m$ that
should be verified by neutron experiments on single crystals.

As follows from Table~\ref{T-QMC}, different values of $J_1'/J_1$ yield rather
similar $T_{\rm N}$ and $m$. This surprisingly robust behavior
originates from peculiar properties of spin-3/2 alternating Heisenberg chains
that show low spin gaps in the $0.41<J_1'/J_1<1$ range between the critical
points at $J_1'/J_1=0.41$ and $J_1'/J_1=1$, where the spin gap vanishes (see
Sec.~\ref{S-quant-class} and Ref.~\onlinecite{yajima1996}). To evaluate the
optimal $J_1'/J_1$, we trace the general trends.  First, smaller $J_1'/J_1$
ratios yield better agreement with the experimental $T_{\rm N}$. Second,
$m$ is nearly independent of $J_1'/J_1$ for $J_1'/J_1\!\geq\!0.5$.
Thus, the parametrization $J_1:J_1':\jici:\jicii\!\simeq\!1:0.5:-0.02:0.02$
with $J_1\!\simeq\!50$\,K is the optimal choice, since it conforms to the
experimental $\chi(T)$ dependence, accurately reproduces the experimental
$T_{\rm N}$ as well as the $g$-factor, and is in reasonable agreement with the
experimental $m$.

\section{\label{S-quant-class}Quantum--classical crossover}
For a correct application of classical models, it is crucial to know at which point
quantum effects become relevant and the classical
approximation breaks down.  In this respect, the quasi-1D magnetic model
of \crbpo\ is a promising candidate. The leading exchange couplings
$J_1$ and $J_1'$ form alternating chains, while the interchain couplings
are substantially smaller.  Besides, \crbpo\ lacks frustration, thus
quantum MC simulations can be performed.

First, we study nearest-neighbor spin correlations using QMC and classical MC
simulations. The difference between the two results is directly related to
quantum fluctuations: while QMC correctly accounts for the quantum behavior at
finite temperatures, the classical MC method captures thermal fluctuations,
only. 

In the following, we consider a 1D $J_1-J_1'$ magnetic model, where the interchain
couplings are neglected completely. For $J_1'/J_1$\,=\,0.5, we simulate
the temperature dependence of the diagonal spin correlations
$\langle{}S^z_iS^z_{i+1}\rangle$ and
$\langle{}S^z_{i+1}S^z_{i+2}\rangle$, where $i$\,=\,0 corresponds to
the stronger coupling $J_1$, while $i$\,=\,1 corresponds to $J_1'$. 

\begin{figure}[tbp]
\includegraphics[width=8.6cm]{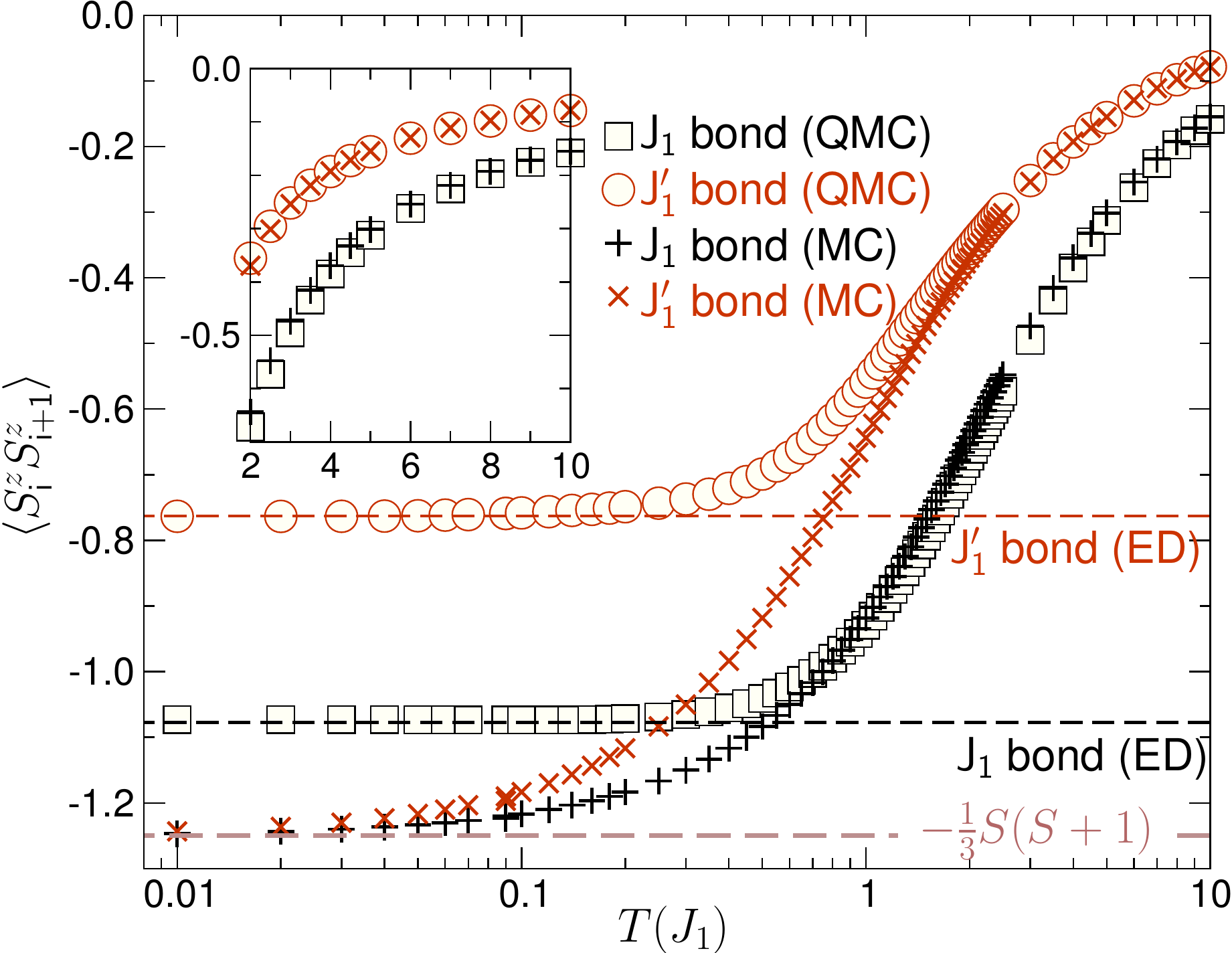}
\caption{\label{F-Sz}(Color online) Temperature dependence of the
nearest-neighbor diagonal spin correlations $\langle{}S^z_iS^z_{i+1}\rangle$ in
the $S$\,=\,3/2 alternating Heisenberg chain model ($J_1'$/$J_1$\,=\,0.5).  QMC
results are plotted with squares ($J_1$) and circles ($J_1'$), the classical MC
results are depicted with pluses ($J_1$) and crosses ($J_1'$). The dashed lines
denote the ED results for the quantum model.  The $\frac{1}{3}S(S+1)$ line
marks the maximal diagonal correlation for $S$\,=\,3/2 spins.  Inset: at high
temperatures ($T$\,$>$\,$2J_1$), the QMC and MC curves become almost
indistinguishable.
} \end{figure}

The resulting curves are plotted in Fig.~\ref{F-Sz}. In the high-temperature
range (see inset in Fig.~\ref{F-Sz}), the classical and quantum results are
practically indistinguishable. At $T$\,$\simeq$\,$J_1$, the curves start to
deviate significantly.  The classical MC curves exhibit an asymptotic behavior
and join at $T$\,=\,0. This is in line with a complete elimination of thermal
fluctuations, thereby the spin correlations reach their extremal value for
$S$\,=\,3/2 spins, which amounts to $-[S(S+1)]/3$\,=\,$-1.25$. 
	
Although the QMC curves exhibit qualitatively similar behavior at low
temperatures, they saturate at substantially higher values of
$\langle{}S^z_{i}S^z_{i+1}\rangle$. Dissimilar to the classical model, the
$i$\,=\,0 ($J_1$) and $i$\,=\,1 ($J_1'$) QMC curves saturate at different
values of $\langle{}S^z_{i}S^z_{i+1}\rangle$, indicating the onset of
dimerization.

In the vicinity of the broad maximum of the magnetic susceptibility ($T^{\rm
max}$\,$\simeq$\,2\,$J_1$), and even at lower temperatures, the classical MC
simulations are in excellent agreement with the quantum model. Therefore,
experimental thermodynamic data $[$e.g., $\chi(T)$ and magnetic specific heat$]$
for $T$\,$\geq$\,$J_1$ should be well reproduced within the classical
approximation. This empirical rule follows earlier experimental and numerical
results for uniform spin-3/2 chains.\cite{CsVCl3_INS_1995,
*CsVCl3_CsVBr3_INS_1999} It can also be generalized to quasi-2D and 3D systems,
owing to further suppression of quantum fluctuations in higher dimensions.
However, in 0D systems, such as isolated dimers, quantum effects are expected
to play a more substantial role.

To explore the evolution of quantum fluctuations by a gradual crossover
from 1D to 0D regime, we evaluate the GS diagonal spin correlations for
different ratios $J_1'$/$J_1$. Although QMC simulations are not applicable at
$T$\,=\,0, the temperature evolution of spin correlations in
Fig.~\ref{F-Sz} suggests only a marginal change in
$\langle{}S^z_{i}S^z_{i+1}\rangle$ below 0.03\,$J_1$.  Thus,
finite-temperature spin correlations simulated at $T$\,=\,0.03\,$J_1$
can be regarded as a rather accurate measure of GS ($T$\,=\,0) spin
correlations.

\begin{figure}[tbp]
\includegraphics[width=8.6cm]{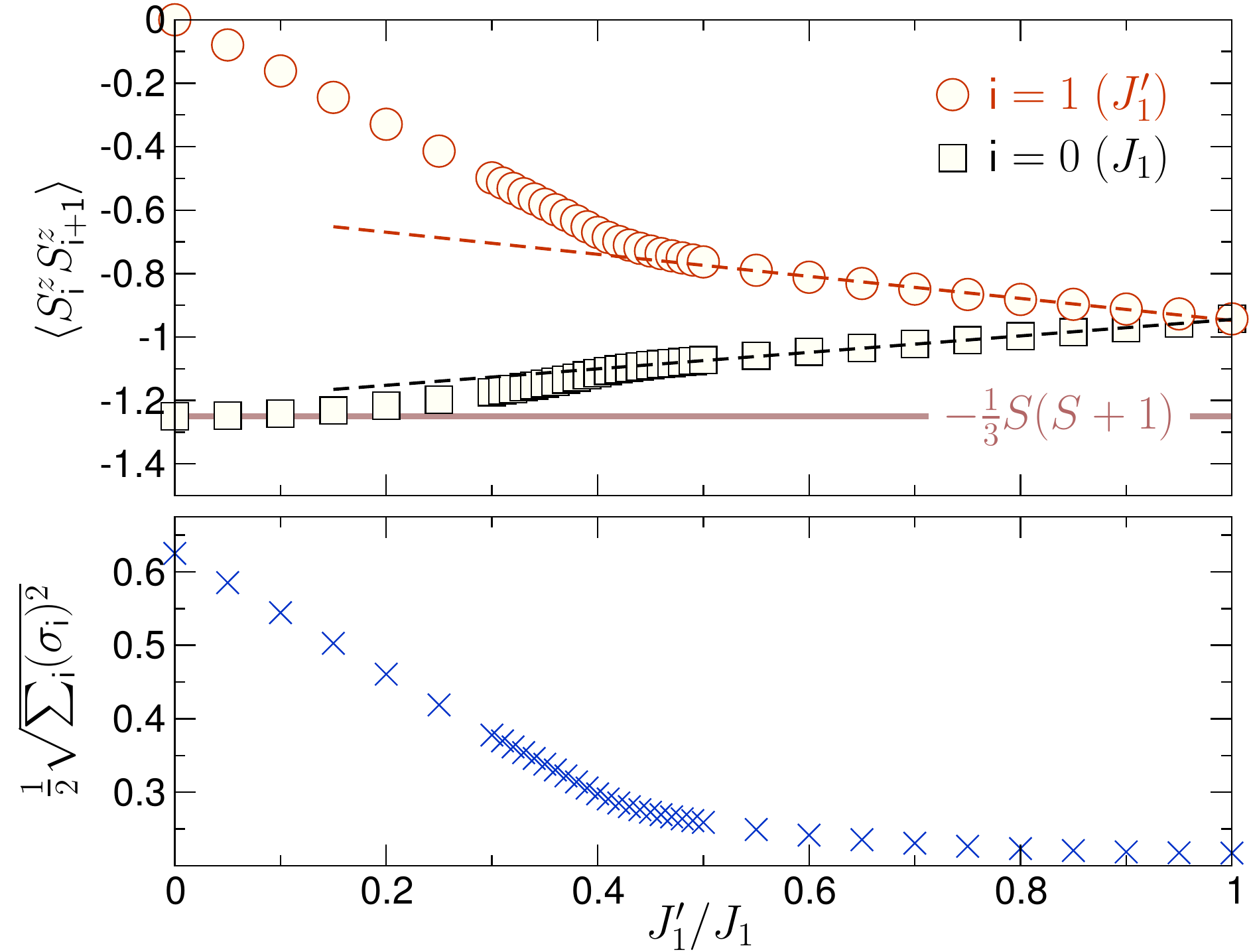}
\caption{\label{F-Sz-ahc}(Color online) Top: nearest-neighbor diagonal spin
correlations $\langle{}S^z_iS^z_{i+1}\rangle$ in the $S$\,=\,3/2 alternating
Heisenberg chain model as a function of the alternation ratio $J_1'$/$J_1$ at
$T$\,=\,0.03\,$J_1$.  Dashed lines are guides to the eye.  Bottom: root mean
square of
$\sigma_i$\,$\equiv$\,$\left(\langle{}S^z_iS^z_{i+1}\rangle-\frac{1}{3}S[S+1]\right)$
($i$\,=\,1, 2), reflecting the discrepancy between the quantum and the
classical models.} 
\end{figure}

Fig.~\ref{F-Sz-ahc} (top) shows the resulting
$\langle{}S^z_iS^z_{i+1}\rangle$($J_1'$/$J_1$) dependencies. In the
uniform-chain limit ($J_1'/J_1\!=\!1$), the correlations for $i\!=\!0$ and 1
coincide. Dimerization gives rise to a quasilinear behavior observed down to
$J_1'/J_1\!\simeq\!0.4$. At this point, both curves exhibit a pronounced kink,
and their further evolution is different: the $i\!=\!0$ ($J_1$) curve rapidly drops
almost down to the extremal value $\frac{1}{3}S[S+1]$, while the $i\!=\!1$
($J_1'$) curve exhibits a quasilinear growth up to zero correlation in the
$J_1'/J_1\!=\!0$ limit. This behavior is remarkably different from the
classical case, where the constant spin-spin correlation of $-\frac13S(S+1)$ is
expected.

To sharpen the crossover at $J_1'$/$J_1$\,$\simeq$\,0.4, we consider root mean
square deviation for
$\sigma_i\!\equiv\!\left(\langle{}S^z_iS^z_{i+1}\rangle-\frac{1}{3}S(S+1)\right)$
($i\!=\!1,2$), which reflects the difference between the quantum and classical
models. The resulting dependence is shown in Fig.~\ref{F-Sz-ahc} (bottom). For
$0.5\!\leq\!J_1'/J_1\!\leq\!1$, only a marginal increase is observed, while at
$J_1'/J_1\!\simeq\!0.4$ the slope changes distinctly. For $J_1'/J_1\!<\!0.4$, a
quasilinear behavior is restored.

The drastic change in the spin-spin correlations around $J_1'/J_1\!\simeq\!0.4$
indicates the critical point of bond-alternating spin-3/2 chains. While such
chains generally have gapped GS, they become gapless in the
uniform-chain limit ($J_1'/J_1\!=\!1$) and at the critical point of
$J_1'/J_1\!\simeq\!0.41$, according to density-matrix renormalization
group\cite{yajima1996} and QMC studies.\cite{yamamoto1997} Qualitatively, the
critical point at $J_1'/J_1\!\simeq\!0.41$ can be understood as the transition
between two different valence-bond-solid (VBS)-type GSs. 

The VBS state entails singlet pairs formed by individual spin-1/2 entities
comprising spin-3/2 sites of the lattice.\cite{affleck1987,*affleck1988}
Although it is not an exact GS of the Heisenberg Hamiltonian in
Eq.~\eqref{E-H}, the references to relevant VBS states are instrumental in
understanding peculiarities of the magnetic behavior. The spin-3/2 chain can form two
VBS states, the (3,0) state having three singlet pairs on one bond and no pairs
on the contiguous bonds, or the (2,1) state having two singlet pairs on one
bond and one singlet pair on each of the neighboring bonds. At low $J_1'/J_1$,
alternating spin-3/2 chains are close to the (3,0) regime, as shown by the
drastic difference between the spin-spin correlations for $J_1$ and $J_1'$. At
$J_1'/J_1>0.41$, the system resembles the (2,1) VBS\cite{yamamoto1997} and
eventually develops the GS of a uniform spin-3/2 chain with equal correlations
on the $J_1$ and $J_1'$ bonds (Fig.~\ref{F-Sz-ahc}). Note that the
low-temperature physics of alternating spin-3/2 chains is essentially governed
by quantum effects and can not be reproduced in the classical approximation.

\section{\label{S-sum}Discussion and Summary}
The formation of alternating spin chains contrasts with the presence of
isolated Cr$_2$O$_9$ dimers in the crystal structure of \crbpo. However, such
differences between the structure and the ensuing spin lattice are not unusual,
especially for phosphates, vanadates, and other compounds with complex anions that
are capable of mediating the superexchange. For example, a large superexchange
coupling via the PO$_4$ tetrahedra between the structural dimers is a salient
feature of the spin-1/2 alternating-chain magnet (VO)$_2[$P$_2$O$_7]$
(Ref.~\onlinecite{garrett1997}). In spin-1/2 V$^{4+}$ and Cu$^{2+}$ compounds,
interdimer couplings often exceed the intradimer exchange, whereas the latter
is small or even negligible, as in Cu$_2[$P$_2$O$_7]$
(Ref.~\onlinecite{janson2011}) or VO$[$HPO$_4]\cdot \frac12$H$_2$O
(Ref.~\onlinecite{tennant1997}). 

The \crbpo\ case is different, though, because the intradimer coupling $J_1$
exceeds the interdimer coupling $J_1'$. The reason behind the sizable
intradimer coupling is the strong direct exchange that was previously observed
in LiCrO$_2$.\cite{LiCrO2_DFT_2007} While the Cr--O--Cr angles amount to
$87.8^{\circ}$ and impede the AF superexchange, the direct overlap of Cr $d$
orbitals is strong enough to facilitate the sizable AF $J_1$. This mechanism is
not operative in typical spin-1/2 systems with structural dimers, because
V$^{4+}$ cations form VO$_5$ square pyramids and remain on opposite sides of
the basal plane of such pyramids, thus featuring only a weak direct overlap of
the magnetic $d_{xy}$ orbitals. In the case of Cu$^{2+}$, the magnetic orbital
has the $d_{x^2-y^2}$ symmetry, unfavorable for the direct overlap.

It is also instructive to compare different interdimer couplings in \crbpo.
Similar to other transition-metal phosphates,\cite{tsirlin2008,janson2011} the
Cr-based Wannier functions show sizable contributions of Cr $3d$ and O $2p$
states, only (Fig.~\ref{F_wf}). Phosphorous states weakly contribute to the
magnetic orbitals and play a minor role in the superexchange running via the
Cr--O...O--Cr pathways. The efficiency of these pathways intimately depends on
details of the crystal structure.

Remarkably, \crbpo\ not only entails the spin lattice of alternating chains,
but also features the alternation ratio of $J_1'/J_1\simeq 0.5$, which is close
to the critical point with the gapless ground state at $J_1'/J_1\simeq 0.41$. The
proximity to this critical point is one of the reasons behind the long-range AF
ordering and the sizable N\'eel temperature in \crbpo. Spin-1/2 systems with
the larger $J_1'/J_1\!\simeq\!0.6$ (i.e., weaker dimerization) and comparable
interchain couplings\footnote{The interchain couplings should be considered
irrespective of their sign and together with the relevant coordination numbers.
In \crbpo, the effective interchain coupling is
$J_{\eff}=3J_{\rm ic2}+6|J_{\rm ic1}|=0.18J_1$, i.e., $J_{\rm ic}/J_1=0.09$ for the typical
coordination number of 2 as, e.g., in Pb$_2[$V$_3$O$_9]$.} still have the
spin-gap ground state, as in Pb$_2[$V$_3$O$_9]$ (Ref.~\onlinecite{tsirlin2011c})
and Ag(VO)$[$AsO$_4]$ (Ref.~\onlinecite{tsirlin2011b}). The large number of
interchain couplings per magnetic site (the coordination number amounts to six
and three for $J_{\rm ic1}$ and $J_{\rm ic2}$, respectively), as found in
\crbpo, may also reduce quantum fluctuations. By contrast, an isolated alternating $S\!=\!3/2$ chain
with $J_1'/J_1\simeq 0.5$ would feature an excitation gap of $\sim$0.25\,$J_1$
with no long-range order.\cite{yamamoto1997} 

While the critical point at $J_1'/J_1\simeq 0.41$ has been widely studied
theoretically,\cite{yajima1996,yamamoto1997} experimental probes of this regime
and even experimental examples of spin-3/2 alternating chains have not been
reported. Therefore, \crbpo\ may be interesting as an experimentally available
spin-3/2 alternating-chain system in the vicinity of this critical point.
Although \crbpo\ is not perfectly 1D and features the long-range AF order owing
to non-zero interchain couplings, an experimental study of spin-spin
correlations with, e.g., inelastic neutron scattering, could be instructive.
Moreover, the application of external pressure might change the $J_1'/J_1$
ratio, thus giving access to the peculiar evolution of spin-spin correlations
across the critical point (Fig.~\ref{F-Sz-ahc}). Detailed studies of
the isostructural phase Fe$_2[$BP$_3$O$_{12}]$ (Ref.~\onlinecite{zhang2010}) may be
instructive as well, because this compound should feature a similar spin
lattice in the more classical regime of spin-5/2.

In summary, we extensively characterized the magnetic behavior of \crbpo\ using
magnetic susceptibility, neutron diffraction, and ESR measurements. The
long-range AF order established below $T_{\rm N}\!=\!28$\,K is described with
the propagation vector $\kv=0$, whereby the spins on nearest-neighbor Cr atoms
within the $ab$ plane as well as along the $c$-axis direction are antiparallel.  On
the microscopic level, \crbpo\ features $S\!=\!3/2$ Heisenberg chains with an
alternation of the nearest neighbor couplings $J_1$ and $J_1'$.  The ratio
$J_1'/J_1\simeq0.5$ renders \crbpo\ as an interesting model system lying close
to the critical point at $J_1'/J_1\simeq 0.41$, where the spin-3/2 alternating
chain has the gapless ground state. The chains are coupled by two nonequivalent
interchain exchanges: the FM \jici\ and the AF \jicii, both of the order of
1--2\,K.  The microscopic model is in excellent agreement with the experimental
magnetic structure (see Fig.~\ref{F-str}): the nearest-neighbor spins within a
chain are antiparallel, in accord with the AF nature of $J_1$ and $J_1'$.
Moreover, even the interchain coupling regime is compatible with FM \jici\ and
AF \jicii, thus \crbpo\ lacks any appreciable frustration effects.

\section*{Acknowledgements}
We are grateful to Yurii Prots and Horst Borrmann for X-ray diffraction
measurements, and Alim Ormeci for fruitful discussions.  SC, ZJZ, MBT and
JTZ acknowledge the financial support from CAS/SAFEA International Partnership
program for Creative Research Teams (grant number 51121064) and the MPG-CAS
partner group.  A.~A.~T. was supported by the Alexander von Humboldt Foundation
and the Mobilitas grant MTT-77 of the ESF.


\begin{thebibliography}{60}%
\makeatletter
\providecommand \@ifxundefined [1]{%
 \@ifx{#1\undefined}
}%
\providecommand \@ifnum [1]{%
 \ifnum #1\expandafter \@firstoftwo
 \else \expandafter \@secondoftwo
 \fi
}%
\providecommand \@ifx [1]{%
 \ifx #1\expandafter \@firstoftwo
 \else \expandafter \@secondoftwo
 \fi
}%
\providecommand \natexlab [1]{#1}%
\providecommand \enquote  [1]{``#1''}%
\providecommand \bibnamefont  [1]{#1}%
\providecommand \bibfnamefont [1]{#1}%
\providecommand \citenamefont [1]{#1}%
\providecommand \href@noop [0]{\@secondoftwo}%
\providecommand \href [0]{\begingroup \@sanitize@url \@href}%
\providecommand \@href[1]{\@@startlink{#1}\@@href}%
\providecommand \@@href[1]{\endgroup#1\@@endlink}%
\providecommand \@sanitize@url [0]{\catcode `\\12\catcode `\$12\catcode
  `\&12\catcode `\#12\catcode `\^12\catcode `\_12\catcode `\%12\relax}%
\providecommand \@@startlink[1]{}%
\providecommand \@@endlink[0]{}%
\providecommand \url  [0]{\begingroup\@sanitize@url \@url }%
\providecommand \@url [1]{\endgroup\@href {#1}{\urlprefix }}%
\providecommand \urlprefix  [0]{URL }%
\providecommand \Eprint [0]{\href }%
\providecommand \doibase [0]{http://dx.doi.org/}%
\providecommand \selectlanguage [0]{\@gobble}%
\providecommand \bibinfo  [0]{\@secondoftwo}%
\providecommand \bibfield  [0]{\@secondoftwo}%
\providecommand \translation [1]{[#1]}%
\providecommand \BibitemOpen [0]{}%
\providecommand \bibitemStop [0]{}%
\providecommand \bibitemNoStop [0]{.\EOS\space}%
\providecommand \EOS [0]{\spacefactor3000\relax}%
\providecommand \BibitemShut  [1]{\csname bibitem#1\endcsname}%
\let\auto@bib@innerbib\@empty
\bibitem [{\citenamefont {Bethe}(1931)}]{Bethe_Ansatz}%
  \BibitemOpen
  \bibfield  {author} {\bibinfo {author} {\bibfnamefont {H.}~\bibnamefont
  {Bethe}},\ }\href {\doibase 10.1007/BF01341708} {\bibfield  {journal}
  {\bibinfo  {journal} {Z. Phys. A}\ }\textbf {\bibinfo {volume} {71}},\
  \bibinfo {pages} {205} (\bibinfo {year} {1931})}\BibitemShut {NoStop}%
\bibitem [{\citenamefont {Majumdar}\ and\ \citenamefont
  {Ghosh}(1969{\natexlab{a}})}]{MG_1}%
  \BibitemOpen
  \bibfield  {author} {\bibinfo {author} {\bibfnamefont {C.~K.}\ \bibnamefont
  {Majumdar}}\ and\ \bibinfo {author} {\bibfnamefont {D.~K.}\ \bibnamefont
  {Ghosh}},\ }\href {\doibase 10.1063/1.1664978} {\bibfield  {journal}
  {\bibinfo  {journal} {J. Math. Phys.}\ }\textbf {\bibinfo {volume} {10}},\
  \bibinfo {pages} {1388} (\bibinfo {year} {1969}{\natexlab{a}})}\BibitemShut
  {NoStop}%
\bibitem [{\citenamefont {Majumdar}\ and\ \citenamefont
  {Ghosh}(1969{\natexlab{b}})}]{MG_2}%
  \BibitemOpen
  \bibfield  {author} {\bibinfo {author} {\bibfnamefont {C.~K.}\ \bibnamefont
  {Majumdar}}\ and\ \bibinfo {author} {\bibfnamefont {D.~K.}\ \bibnamefont
  {Ghosh}},\ }\href {\doibase 10.1063/1.1664979} {\bibfield  {journal}
  {\bibinfo  {journal} {J. Math. Phys.}\ }\textbf {\bibinfo {volume} {10}},\
  \bibinfo {pages} {1399} (\bibinfo {year} {1969}{\natexlab{b}})}\BibitemShut
  {NoStop}%
\bibitem [{\citenamefont {Shastry}\ and\ \citenamefont
  {Sutherland}(1981)}]{Shastry_Suth}%
  \BibitemOpen
  \bibfield  {author} {\bibinfo {author} {\bibfnamefont {B.~S.}\ \bibnamefont
  {Shastry}}\ and\ \bibinfo {author} {\bibfnamefont {B.}~\bibnamefont
  {Sutherland}},\ }\href {\doibase 10.1016/0378-4363(81)90838-X} {\bibfield
  {journal} {\bibinfo  {journal} {Physica B+C}\ }\textbf {\bibinfo {volume}
  {108}},\ \bibinfo {pages} {1069} (\bibinfo {year} {1981})}\BibitemShut
  {NoStop}%
\bibitem [{\citenamefont {Dingle}\ \emph {et~al.}(1969)\citenamefont {Dingle},
  \citenamefont {Lines},\ and\ \citenamefont {Holt}}]{dingle1969}%
  \BibitemOpen
  \bibfield  {author} {\bibinfo {author} {\bibfnamefont {R.}~\bibnamefont
  {Dingle}}, \bibinfo {author} {\bibfnamefont {M.~E.}\ \bibnamefont {Lines}}, \
  and\ \bibinfo {author} {\bibfnamefont {S.~L.}\ \bibnamefont {Holt}},\ }\href
  {\doibase 10.1103/PhysRev.187.643} {\bibfield  {journal} {\bibinfo  {journal}
  {Phys. Rev.}\ }\textbf {\bibinfo {volume} {187}},\ \bibinfo {pages} {643}
  (\bibinfo {year} {1969})}\BibitemShut {NoStop}%
\bibitem [{\citenamefont {Hutchings}\ \emph {et~al.}(1972)\citenamefont
  {Hutchings}, \citenamefont {Shirane}, \citenamefont {Birgeneau},\ and\
  \citenamefont {Holt}}]{hutchings1972}%
  \BibitemOpen
  \bibfield  {author} {\bibinfo {author} {\bibfnamefont {M.~T.}\ \bibnamefont
  {Hutchings}}, \bibinfo {author} {\bibfnamefont {G.}~\bibnamefont {Shirane}},
  \bibinfo {author} {\bibfnamefont {R.~J.}\ \bibnamefont {Birgeneau}}, \ and\
  \bibinfo {author} {\bibfnamefont {S.~L.}\ \bibnamefont {Holt}},\ }\href
  {\doibase 10.1103/PhysRevB.5.1999} {\bibfield  {journal} {\bibinfo  {journal}
  {Phys. Rev. B}\ }\textbf {\bibinfo {volume} {5}},\ \bibinfo {pages} {1999}
  (\bibinfo {year} {1972})}\BibitemShut {NoStop}%
\bibitem [{\citenamefont {{de Jonge}}\ \emph {et~al.}(1978)\citenamefont {{de
  Jonge}}, \citenamefont {Hijmans}, \citenamefont {Boersma}, \citenamefont
  {Schouten},\ and\ \citenamefont {Kopinga}}]{dejonge1978}%
  \BibitemOpen
  \bibfield  {author} {\bibinfo {author} {\bibfnamefont {W.~J.~M.}\
  \bibnamefont {{de Jonge}}}, \bibinfo {author} {\bibfnamefont {J.~P. A.~M.}\
  \bibnamefont {Hijmans}}, \bibinfo {author} {\bibfnamefont {F.}~\bibnamefont
  {Boersma}}, \bibinfo {author} {\bibfnamefont {J.~C.}\ \bibnamefont
  {Schouten}}, \ and\ \bibinfo {author} {\bibfnamefont {K.}~\bibnamefont
  {Kopinga}},\ }\href {\doibase 10.1103/PhysRevB.17.2922} {\bibfield  {journal}
  {\bibinfo  {journal} {Phys. Rev. B}\ }\textbf {\bibinfo {volume} {17}},\
  \bibinfo {pages} {2922} (\bibinfo {year} {1978})}\BibitemShut {NoStop}%
\bibitem [{\citenamefont {Itoh}\ \emph {et~al.}(1997)\citenamefont {Itoh},
  \citenamefont {Tanaka},\ and\ \citenamefont {Otomo}}]{itoh1997}%
  \BibitemOpen
  \bibfield  {author} {\bibinfo {author} {\bibfnamefont {S.}~\bibnamefont
  {Itoh}}, \bibinfo {author} {\bibfnamefont {H.}~\bibnamefont {Tanaka}}, \ and\
  \bibinfo {author} {\bibfnamefont {T.}~\bibnamefont {Otomo}},\ }\href
  {\doibase 10.1143/JPSJ.66.455} {\bibfield  {journal} {\bibinfo  {journal} {J.
  Phys. Soc. Jpn.}\ }\textbf {\bibinfo {volume} {66}},\ \bibinfo {pages} {455}
  (\bibinfo {year} {1997})}\BibitemShut {NoStop}%
\bibitem [{\citenamefont {Itoh}\ \emph {et~al.}(2002)\citenamefont {Itoh},
  \citenamefont {Tanaka},\ and\ \citenamefont {Bull}}]{itoh2002}%
  \BibitemOpen
  \bibfield  {author} {\bibinfo {author} {\bibfnamefont {S.}~\bibnamefont
  {Itoh}}, \bibinfo {author} {\bibfnamefont {H.}~\bibnamefont {Tanaka}}, \ and\
  \bibinfo {author} {\bibfnamefont {M.~J.}\ \bibnamefont {Bull}},\ }\href
  {\doibase 10.1143/JPSJ.71.1148} {\bibfield  {journal} {\bibinfo  {journal}
  {J. Phys. Soc. Jpn.}\ }\textbf {\bibinfo {volume} {71}},\ \bibinfo {pages}
  {1148} (\bibinfo {year} {2002})}\BibitemShut {NoStop}%
\bibitem [{\citenamefont {Chandra}\ and\ \citenamefont
  {Doucot}(1988)}]{chandra1988}%
  \BibitemOpen
  \bibfield  {author} {\bibinfo {author} {\bibfnamefont {P.}~\bibnamefont
  {Chandra}}\ and\ \bibinfo {author} {\bibfnamefont {B.}~\bibnamefont
  {Doucot}},\ }\href {\doibase 10.1103/PhysRevB.38.9335} {\bibfield  {journal}
  {\bibinfo  {journal} {Phys. Rev. B}\ }\textbf {\bibinfo {volume} {38}},\
  \bibinfo {pages} {9335} (\bibinfo {year} {1988})}\BibitemShut {NoStop}%
\bibitem [{\citenamefont {Dagotto}\ and\ \citenamefont
  {Moreo}(1989)}]{dagotto1989}%
  \BibitemOpen
  \bibfield  {author} {\bibinfo {author} {\bibfnamefont {E.}~\bibnamefont
  {Dagotto}}\ and\ \bibinfo {author} {\bibfnamefont {A.}~\bibnamefont
  {Moreo}},\ }\href {\doibase 10.1103/PhysRevB.39.4744} {\bibfield  {journal}
  {\bibinfo  {journal} {Phys. Rev. B}\ }\textbf {\bibinfo {volume} {39}},\
  \bibinfo {pages} {4744} (\bibinfo {year} {1989})}\BibitemShut {NoStop}%
\bibitem [{\citenamefont {Rousochatzakis}\ \emph {et~al.}(2012)\citenamefont
  {Rousochatzakis}, \citenamefont {L\"auchli},\ and\ \citenamefont
  {Moessner}}]{rousochatzakis2012}%
  \BibitemOpen
  \bibfield  {author} {\bibinfo {author} {\bibfnamefont {I.}~\bibnamefont
  {Rousochatzakis}}, \bibinfo {author} {\bibfnamefont {A.~M.}\ \bibnamefont
  {L\"auchli}}, \ and\ \bibinfo {author} {\bibfnamefont {R.}~\bibnamefont
  {Moessner}},\ }\href {\doibase 10.1103/PhysRevB.85.104415} {\bibfield
  {journal} {\bibinfo  {journal} {Phys. Rev. B}\ }\textbf {\bibinfo {volume}
  {85}},\ \bibinfo {pages} {104415} (\bibinfo {year} {2012})}\BibitemShut
  {NoStop}%
\bibitem [{\citenamefont {Haldane}(1983)}]{haldane1983}%
  \BibitemOpen
  \bibfield  {author} {\bibinfo {author} {\bibfnamefont {F.}~\bibnamefont
  {Haldane}},\ }\href {\doibase 10.1103/PhysRevLett.50.1153} {\bibfield
  {journal} {\bibinfo  {journal} {Phys. Rev. Lett.}\ }\textbf {\bibinfo
  {volume} {50}},\ \bibinfo {pages} {1153} (\bibinfo {year}
  {1983})}\BibitemShut {NoStop}%
\bibitem [{\citenamefont {Itoh}\ \emph {et~al.}(1995)\citenamefont {Itoh},
  \citenamefont {Endoh}, \citenamefont {Kakurai},\ and\ \citenamefont
  {Tanaka}}]{CsVCl3_INS_1995}%
  \BibitemOpen
  \bibfield  {author} {\bibinfo {author} {\bibfnamefont {S.}~\bibnamefont
  {Itoh}}, \bibinfo {author} {\bibfnamefont {Y.}~\bibnamefont {Endoh}},
  \bibinfo {author} {\bibfnamefont {K.}~\bibnamefont {Kakurai}}, \ and\
  \bibinfo {author} {\bibfnamefont {H.}~\bibnamefont {Tanaka}},\ }\href
  {\doibase 10.1103/PhysRevLett.74.2375} {\bibfield  {journal} {\bibinfo
  {journal} {Phys. Rev. Lett.}\ }\textbf {\bibinfo {volume} {74}},\ \bibinfo
  {pages} {2375} (\bibinfo {year} {1995})}\BibitemShut {NoStop}%
\bibitem [{\citenamefont {Itoh}\ \emph {et~al.}(1999)\citenamefont {Itoh},
  \citenamefont {Endoh}, \citenamefont {Kakurai}, \citenamefont {Tanaka},
  \citenamefont {Bennington}, \citenamefont {Perring}, \citenamefont {Ohoyama},
  \citenamefont {Harris}, \citenamefont {Nakajima},\ and\ \citenamefont
  {Frost}}]{CsVCl3_CsVBr3_INS_1999}%
  \BibitemOpen
  \bibfield  {author} {\bibinfo {author} {\bibfnamefont {S.}~\bibnamefont
  {Itoh}}, \bibinfo {author} {\bibfnamefont {Y.}~\bibnamefont {Endoh}},
  \bibinfo {author} {\bibfnamefont {K.}~\bibnamefont {Kakurai}}, \bibinfo
  {author} {\bibfnamefont {H.}~\bibnamefont {Tanaka}}, \bibinfo {author}
  {\bibfnamefont {S.~M.}\ \bibnamefont {Bennington}}, \bibinfo {author}
  {\bibfnamefont {T.~G.}\ \bibnamefont {Perring}}, \bibinfo {author}
  {\bibfnamefont {K.}~\bibnamefont {Ohoyama}}, \bibinfo {author} {\bibfnamefont
  {M.~J.}\ \bibnamefont {Harris}}, \bibinfo {author} {\bibfnamefont
  {K.}~\bibnamefont {Nakajima}}, \ and\ \bibinfo {author} {\bibfnamefont
  {C.~D.}\ \bibnamefont {Frost}},\ }\href {\doibase 10.1103/PhysRevB.59.14406}
  {\bibfield  {journal} {\bibinfo  {journal} {Phys. Rev. B}\ }\textbf {\bibinfo
  {volume} {59}},\ \bibinfo {pages} {14406} (\bibinfo {year}
  {1999})}\BibitemShut {NoStop}%
\bibitem [{\citenamefont {Payen}\ \emph {et~al.}(1992)\citenamefont {Payen},
  \citenamefont {Mutka}, \citenamefont {Soubeyroux}, \citenamefont
  {Molini\'e},\ and\ \citenamefont {Colombet}}]{AgCrP2S6_1992}%
  \BibitemOpen
  \bibfield  {author} {\bibinfo {author} {\bibfnamefont {C.}~\bibnamefont
  {Payen}}, \bibinfo {author} {\bibfnamefont {H.}~\bibnamefont {Mutka}},
  \bibinfo {author} {\bibfnamefont {J.}~\bibnamefont {Soubeyroux}}, \bibinfo
  {author} {\bibfnamefont {P.}~\bibnamefont {Molini\'e}}, \ and\ \bibinfo
  {author} {\bibfnamefont {P.}~\bibnamefont {Colombet}},\ }\href {\doibase
  10.1016/0304-8853(92)90364-T} {\bibfield  {journal} {\bibinfo  {journal} {J.
  Magn. Magn. Mater.}\ }\textbf {\bibinfo {volume} {104-107}},\ \bibinfo
  {pages} {797} (\bibinfo {year} {1992})}\BibitemShut {NoStop}%
\bibitem [{\citenamefont {Mutka}\ \emph {et~al.}(1993)\citenamefont {Mutka},
  \citenamefont {Payen},\ and\ \citenamefont {Molinie}}]{AgCrP2S6_INS_1993}%
  \BibitemOpen
  \bibfield  {author} {\bibinfo {author} {\bibfnamefont {H.}~\bibnamefont
  {Mutka}}, \bibinfo {author} {\bibfnamefont {C.}~\bibnamefont {Payen}}, \ and\
  \bibinfo {author} {\bibfnamefont {P.}~\bibnamefont {Molinie}},\ }\href
  {\doibase 10.1209/0295-5075/21/5/020} {\bibfield  {journal} {\bibinfo
  {journal} {Europhys. Lett.}\ }\textbf {\bibinfo {volume} {21}},\ \bibinfo
  {pages} {623} (\bibinfo {year} {1993})}\BibitemShut {NoStop}%
\bibitem [{\citenamefont {Pechini}(1967)}]{pechini}%
  \BibitemOpen
  \bibfield  {author} {\bibinfo {author} {\bibfnamefont {M.~P.}\ \bibnamefont
  {Pechini}},\ }\href@noop {} {} (\bibinfo {year} {1967}),\ \bibinfo {note}
  {{U.S.} patent 3.330.397}\BibitemShut {NoStop}%
\bibitem [{\citenamefont {Mi}\ \emph {et~al.}(2000)\citenamefont {Mi},
  \citenamefont {Zhao}, \citenamefont {Mao}, \citenamefont {Huang},
  \citenamefont {Engelhardt},\ and\ \citenamefont {Kniep}}]{crbpo_str_2000}%
  \BibitemOpen
  \bibfield  {author} {\bibinfo {author} {\bibfnamefont {J.-X.}\ \bibnamefont
  {Mi}}, \bibinfo {author} {\bibfnamefont {J.-T.}\ \bibnamefont {Zhao}},
  \bibinfo {author} {\bibfnamefont {S.-Y.}\ \bibnamefont {Mao}}, \bibinfo
  {author} {\bibfnamefont {Y.-X.}\ \bibnamefont {Huang}}, \bibinfo {author}
  {\bibfnamefont {H.}~\bibnamefont {Engelhardt}}, \ and\ \bibinfo {author}
  {\bibfnamefont {R.}~\bibnamefont {Kniep}},\ }\href@noop {} {\bibfield
  {journal} {\bibinfo  {journal} {Z. Kristallogr. New Cryst. Struct.}\ }\textbf
  {\bibinfo {volume} {215}},\ \bibinfo {pages} {201} (\bibinfo {year}
  {2000})}\BibitemShut {NoStop}%
\bibitem [{Note1()}]{Note1}%
  \BibitemOpen
  \bibinfo {note} {The $^{11}$B-containing chemical (H$_3^{11}$BO$_3$) has been
  used to prepare a sample for neutron powder diffraction.}\BibitemShut {Stop}%
\bibitem [{\citenamefont {Larson}\ and\ \citenamefont {von
  Dreele}(1994)}]{GSAS}%
  \BibitemOpen
  \bibfield  {author} {\bibinfo {author} {\bibfnamefont {A.}~\bibnamefont
  {Larson}}\ and\ \bibinfo {author} {\bibfnamefont {R.}~\bibnamefont {von
  Dreele}},\ }\href@noop {} {\enquote {\bibinfo {title} {General structure
  analysis system {\textsc{(gsas)}}},}\ }\bibinfo {howpublished} {Los Alamos
  National Laboratory Report LAUR 86-748} (\bibinfo {year} {1994})\BibitemShut
  {NoStop}%
\bibitem [{\citenamefont {Rodr{\'\i}guez-Carvajal}(1993)}]{fullprof}%
  \BibitemOpen
  \bibfield  {author} {\bibinfo {author} {\bibfnamefont {J.}~\bibnamefont
  {Rodr{\'\i}guez-Carvajal}},\ }\href {\doibase 10.1016/0921-4526(93)90108-I}
  {\bibfield  {journal} {\bibinfo  {journal} {Physica B}\ }\textbf {\bibinfo
  {volume} {192}},\ \bibinfo {pages} {55} (\bibinfo {year} {1993})}\BibitemShut
  {NoStop}%
\bibitem [{\citenamefont {Koepernik}\ and\ \citenamefont
  {Eschrig}(1999)}]{FPLO}%
  \BibitemOpen
  \bibfield  {author} {\bibinfo {author} {\bibfnamefont {K.}~\bibnamefont
  {Koepernik}}\ and\ \bibinfo {author} {\bibfnamefont {H.}~\bibnamefont
  {Eschrig}},\ }\href {\doibase 10.1103/PhysRevB.59.1743} {\bibfield  {journal}
  {\bibinfo  {journal} {Phys. Rev. B}\ }\textbf {\bibinfo {volume} {59}},\
  \bibinfo {pages} {1743} (\bibinfo {year} {1999})}\BibitemShut {NoStop}%
\bibitem [{\citenamefont {Kresse}\ and\ \citenamefont
  {Furthm{\"{u}}ller}(1996{\natexlab{a}})}]{VASP}%
  \BibitemOpen
  \bibfield  {author} {\bibinfo {author} {\bibfnamefont {G.}~\bibnamefont
  {Kresse}}\ and\ \bibinfo {author} {\bibfnamefont {J.}~\bibnamefont
  {Furthm{\"{u}}ller}},\ }\href {\doibase 10.1103/PhysRevB.54.11169} {\bibfield
   {journal} {\bibinfo  {journal} {Phys. Rev. B}\ }\textbf {\bibinfo {volume}
  {54}},\ \bibinfo {pages} {11169} (\bibinfo {year}
  {1996}{\natexlab{a}})}\BibitemShut {NoStop}%
\bibitem [{\citenamefont {Kresse}\ and\ \citenamefont
  {Furthm{\"{u}}ller}(1996{\natexlab{b}})}]{VASP_2}%
  \BibitemOpen
  \bibfield  {author} {\bibinfo {author} {\bibfnamefont {G.}~\bibnamefont
  {Kresse}}\ and\ \bibinfo {author} {\bibfnamefont {J.}~\bibnamefont
  {Furthm{\"{u}}ller}},\ }\href {\doibase 10.1016/0927-0256(96)00008-0}
  {\bibfield  {journal} {\bibinfo  {journal} {Comput. Mater. Sci.}\ }\textbf
  {\bibinfo {volume} {6}},\ \bibinfo {pages} {15} (\bibinfo {year}
  {1996}{\natexlab{b}})}\BibitemShut {NoStop}%
\bibitem [{\citenamefont {Perdew}\ and\ \citenamefont {Wang}(1992)}]{PW92}%
  \BibitemOpen
  \bibfield  {author} {\bibinfo {author} {\bibfnamefont {J.~P.}\ \bibnamefont
  {Perdew}}\ and\ \bibinfo {author} {\bibfnamefont {Y.}~\bibnamefont {Wang}},\
  }\href {\doibase 10.1103/PhysRevB.45.13244} {\bibfield  {journal} {\bibinfo
  {journal} {Phys. Rev. B}\ }\textbf {\bibinfo {volume} {45}},\ \bibinfo
  {pages} {13244} (\bibinfo {year} {1992})}\BibitemShut {NoStop}%
\bibitem [{\citenamefont {Perdew}\ \emph {et~al.}(1996)\citenamefont {Perdew},
  \citenamefont {Burke},\ and\ \citenamefont {Ernzerhof}}]{PBE96}%
  \BibitemOpen
  \bibfield  {author} {\bibinfo {author} {\bibfnamefont {J.~P.}\ \bibnamefont
  {Perdew}}, \bibinfo {author} {\bibfnamefont {K.}~\bibnamefont {Burke}}, \
  and\ \bibinfo {author} {\bibfnamefont {M.}~\bibnamefont {Ernzerhof}},\ }\href
  {\doibase 10.1103/PhysRevLett.77.3865} {\bibfield  {journal} {\bibinfo
  {journal} {Phys. Rev. Lett.}\ }\textbf {\bibinfo {volume} {77}},\ \bibinfo
  {pages} {3865} (\bibinfo {year} {1996})}\BibitemShut {NoStop}%
\bibitem [{\citenamefont {Heyd}\ \emph {et~al.}(2003)\citenamefont {Heyd},
  \citenamefont {Scuseria},\ and\ \citenamefont {Ernzerhof}}]{HSE03}%
  \BibitemOpen
  \bibfield  {author} {\bibinfo {author} {\bibfnamefont {J.}~\bibnamefont
  {Heyd}}, \bibinfo {author} {\bibfnamefont {G.~E.}\ \bibnamefont {Scuseria}},
  \ and\ \bibinfo {author} {\bibfnamefont {M.}~\bibnamefont {Ernzerhof}},\
  }\href {\doibase 10.1063/1.1564060} {\bibfield  {journal} {\bibinfo
  {journal} {J. Chem. Phys.}\ }\textbf {\bibinfo {volume} {118}},\ \bibinfo
  {pages} {8207} (\bibinfo {year} {2003})}\BibitemShut {NoStop}%
\bibitem [{\citenamefont {Heyd}\ and\ \citenamefont {Scuseria}(2004)}]{HSE04}%
  \BibitemOpen
  \bibfield  {author} {\bibinfo {author} {\bibfnamefont {J.}~\bibnamefont
  {Heyd}}\ and\ \bibinfo {author} {\bibfnamefont {G.~E.}\ \bibnamefont
  {Scuseria}},\ }\href {\doibase 10.1063/1.1760074} {\bibfield  {journal}
  {\bibinfo  {journal} {J. Chem. Phys.}\ }\textbf {\bibinfo {volume} {121}},\
  \bibinfo {pages} {1187} (\bibinfo {year} {2004})}\BibitemShut {NoStop}%
\bibitem [{Note2()}]{Note2}%
  \BibitemOpen
  \bibinfo {note} {Large computational demands incurred by the hybrid
  functionals lead to severe restrictions on the number of $k$ points.
  Nevertheless, the results obtained on a sparse $k$ mesh are sufficiently
  accurate and well-converged, owing to the insulating nature of
  Cr$_2[$BP$_3$O$_{12}]$.}\BibitemShut {Stop}%
\bibitem [{\citenamefont {Albuquerque}\ \emph {et~al.}(2007)\citenamefont
  {Albuquerque}, \citenamefont {Alet}, \citenamefont {Corboz}, \citenamefont
  {Dayal}, \citenamefont {Feiguin}, \citenamefont {Fuchs}, \citenamefont
  {Gamper}, \citenamefont {Gull}, \citenamefont {G\"urtler}, \citenamefont
  {Honecker}, \citenamefont {Igarashi}, \citenamefont {K\"orner}, \citenamefont
  {Kozhevnikov}, \citenamefont {L\"auchli}, \citenamefont {Manmana},
  \citenamefont {Matsumoto}, \citenamefont {McCulloch}, \citenamefont {Michel},
  \citenamefont {Noack}, \citenamefont {Pawlowski}, \citenamefont {Pollet},
  \citenamefont {Pruschke}, \citenamefont {Schollw\"ock}, \citenamefont {Todo},
  \citenamefont {Trebst}, \citenamefont {Troyer}, \citenamefont {Werner},\ and\
  \citenamefont {Wessel}}]{ALPS}%
  \BibitemOpen
  \bibfield  {author} {\bibinfo {author} {\bibfnamefont {A.}~\bibnamefont
  {Albuquerque}}, \bibinfo {author} {\bibfnamefont {F.}~\bibnamefont {Alet}},
  \bibinfo {author} {\bibfnamefont {P.}~\bibnamefont {Corboz}}, \bibinfo
  {author} {\bibfnamefont {P.}~\bibnamefont {Dayal}}, \bibinfo {author}
  {\bibfnamefont {A.}~\bibnamefont {Feiguin}}, \bibinfo {author} {\bibfnamefont
  {S.}~\bibnamefont {Fuchs}}, \bibinfo {author} {\bibfnamefont
  {L.}~\bibnamefont {Gamper}}, \bibinfo {author} {\bibfnamefont
  {E.}~\bibnamefont {Gull}}, \bibinfo {author} {\bibfnamefont {S.}~\bibnamefont
  {G\"urtler}}, \bibinfo {author} {\bibfnamefont {A.}~\bibnamefont {Honecker}},
  \bibinfo {author} {\bibfnamefont {R.}~\bibnamefont {Igarashi}}, \bibinfo
  {author} {\bibfnamefont {M.}~\bibnamefont {K\"orner}}, \bibinfo {author}
  {\bibfnamefont {A.}~\bibnamefont {Kozhevnikov}}, \bibinfo {author}
  {\bibfnamefont {A.}~\bibnamefont {L\"auchli}}, \bibinfo {author}
  {\bibfnamefont {S.~R.}\ \bibnamefont {Manmana}}, \bibinfo {author}
  {\bibfnamefont {M.}~\bibnamefont {Matsumoto}}, \bibinfo {author}
  {\bibfnamefont {I.~P.}\ \bibnamefont {McCulloch}}, \bibinfo {author}
  {\bibfnamefont {F.}~\bibnamefont {Michel}}, \bibinfo {author} {\bibfnamefont
  {R.~M.}\ \bibnamefont {Noack}}, \bibinfo {author} {\bibfnamefont
  {G.}~\bibnamefont {Pawlowski}}, \bibinfo {author} {\bibfnamefont
  {L.}~\bibnamefont {Pollet}}, \bibinfo {author} {\bibfnamefont
  {T.}~\bibnamefont {Pruschke}}, \bibinfo {author} {\bibfnamefont
  {U.}~\bibnamefont {Schollw\"ock}}, \bibinfo {author} {\bibfnamefont
  {S.}~\bibnamefont {Todo}}, \bibinfo {author} {\bibfnamefont {S.}~\bibnamefont
  {Trebst}}, \bibinfo {author} {\bibfnamefont {M.}~\bibnamefont {Troyer}},
  \bibinfo {author} {\bibfnamefont {P.}~\bibnamefont {Werner}}, \ and\ \bibinfo
  {author} {\bibfnamefont {S.}~\bibnamefont {Wessel}},\ }\href {\doibase
  10.1016/j.jmmm.2006.10.304} {\bibfield  {journal} {\bibinfo  {journal} {J.
  Magn. Magn. Mater.}\ }\textbf {\bibinfo {volume} {310}},\ \bibinfo {pages}
  {1187} (\bibinfo {year} {2007})},\ \Eprint
  {http://arxiv.org/abs/arXiv:0801.1765} {arXiv:0801.1765} \BibitemShut
  {NoStop}%
\bibitem [{\citenamefont {Zhang}\ \emph {et~al.}(2010)\citenamefont {Zhang},
  \citenamefont {Lin}, \citenamefont {Geng}, \citenamefont {Li}, \citenamefont
  {Zhang}, \citenamefont {He},\ and\ \citenamefont {Cheng}}]{zhang2010}%
  \BibitemOpen
  \bibfield  {author} {\bibinfo {author} {\bibfnamefont {W.-L.}\ \bibnamefont
  {Zhang}}, \bibinfo {author} {\bibfnamefont {C.-S.}\ \bibnamefont {Lin}},
  \bibinfo {author} {\bibfnamefont {L.}~\bibnamefont {Geng}}, \bibinfo {author}
  {\bibfnamefont {Y.-Y.}\ \bibnamefont {Li}}, \bibinfo {author} {\bibfnamefont
  {H.}~\bibnamefont {Zhang}}, \bibinfo {author} {\bibfnamefont {Z.-Z.}\
  \bibnamefont {He}}, \ and\ \bibinfo {author} {\bibfnamefont {W.-D.}\
  \bibnamefont {Cheng}},\ }\href {\doibase 10.1016/j.jssc.2010.03.020}
  {\bibfield  {journal} {\bibinfo  {journal} {J. Solid State Chem.}\ }\textbf
  {\bibinfo {volume} {183}},\ \bibinfo {pages} {1108} (\bibinfo {year}
  {2010})}\BibitemShut {NoStop}%
\bibitem [{\citenamefont {Li}\ \emph {et~al.}(2010)\citenamefont {Li},
  \citenamefont {Zhang},\ and\ \citenamefont {Zhang}}]{li2010}%
  \BibitemOpen
  \bibfield  {author} {\bibinfo {author} {\bibfnamefont {F.~F.}\ \bibnamefont
  {Li}}, \bibinfo {author} {\bibfnamefont {H.~J.}\ \bibnamefont {Zhang}}, \
  and\ \bibinfo {author} {\bibfnamefont {L.~N.}\ \bibnamefont {Zhang}},\ }\href
  {\doibase 10.1107/S1600536810029818} {\bibfield  {journal} {\bibinfo
  {journal} {Acta Cryst.}\ }\textbf {\bibinfo {volume} {E66}},\ \bibinfo
  {pages} {i63} (\bibinfo {year} {2010})}\BibitemShut {NoStop}%
\bibitem [{\citenamefont {Campostrini}\ \emph {et~al.}(2002)\citenamefont
  {Campostrini}, \citenamefont {Hasenbusch}, \citenamefont {Pelissetto},
  \citenamefont {Rossi},\ and\ \citenamefont {Vicari}}]{campostrini2002}%
  \BibitemOpen
  \bibfield  {author} {\bibinfo {author} {\bibfnamefont {M.}~\bibnamefont
  {Campostrini}}, \bibinfo {author} {\bibfnamefont {M.}~\bibnamefont
  {Hasenbusch}}, \bibinfo {author} {\bibfnamefont {A.}~\bibnamefont
  {Pelissetto}}, \bibinfo {author} {\bibfnamefont {P.}~\bibnamefont {Rossi}}, \
  and\ \bibinfo {author} {\bibfnamefont {E.}~\bibnamefont {Vicari}},\ }\href
  {\doibase 10.1103/PhysRevB.65.144520} {\bibfield  {journal} {\bibinfo
  {journal} {Phys. Rev. B}\ }\textbf {\bibinfo {volume} {65}},\ \bibinfo
  {pages} {144520} (\bibinfo {year} {2002})},\ \Eprint
  {http://arxiv.org/abs/cond-mat/0110336} {cond-mat/0110336} \BibitemShut
  {NoStop}%
\bibitem [{\citenamefont {Pelissetto}\ and\ \citenamefont
  {Vicari}(2002)}]{pelissetto2002}%
  \BibitemOpen
  \bibfield  {author} {\bibinfo {author} {\bibfnamefont {A.}~\bibnamefont
  {Pelissetto}}\ and\ \bibinfo {author} {\bibfnamefont {E.}~\bibnamefont
  {Vicari}},\ }\href {\doibase 10.1016/S0370-1573(02)00219-3} {\bibfield
  {journal} {\bibinfo  {journal} {Phys. Rep.}\ }\textbf {\bibinfo {volume}
  {368}},\ \bibinfo {pages} {549} (\bibinfo {year} {2002})},\ \Eprint
  {http://arxiv.org/abs/cond-mat/0012164} {cond-mat/0012164} \BibitemShut
  {NoStop}%
\bibitem [{sup()}]{supplement}%
  \BibitemOpen
  \href@noop {} {}\bibinfo {note} {See Supplementary information for the
  refined neutron diffraction patterns, for the crystal structure refinement,
  ESR data, low-temperature magnetization data, as well as LSDA and GGA density
  of states for the ferromagnetic solution.}\BibitemShut {Stop}%
\bibitem [{\citenamefont {Garrett}\ \emph {et~al.}(1997)\citenamefont
  {Garrett}, \citenamefont {Nagler}, \citenamefont {Tennant}, \citenamefont
  {Sales},\ and\ \citenamefont {Barnes}}]{garrett1997}%
  \BibitemOpen
  \bibfield  {author} {\bibinfo {author} {\bibfnamefont {A.~W.}\ \bibnamefont
  {Garrett}}, \bibinfo {author} {\bibfnamefont {S.~E.}\ \bibnamefont {Nagler}},
  \bibinfo {author} {\bibfnamefont {D.~A.}\ \bibnamefont {Tennant}}, \bibinfo
  {author} {\bibfnamefont {B.~C.}\ \bibnamefont {Sales}}, \ and\ \bibinfo
  {author} {\bibfnamefont {T.}~\bibnamefont {Barnes}},\ }\href {\doibase
  10.1103/PhysRevLett.79.745} {\bibfield  {journal} {\bibinfo  {journal} {Phys.
  Rev. Lett.}\ }\textbf {\bibinfo {volume} {79}},\ \bibinfo {pages} {745}
  (\bibinfo {year} {1997})}\BibitemShut {NoStop}%
\bibitem [{\citenamefont {Tsirlin}\ \emph {et~al.}(2008)\citenamefont
  {Tsirlin}, \citenamefont {Nath}, \citenamefont {Geibel},\ and\ \citenamefont
  {Rosner}}]{tsirlin2008}%
  \BibitemOpen
  \bibfield  {author} {\bibinfo {author} {\bibfnamefont {A.~A.}\ \bibnamefont
  {Tsirlin}}, \bibinfo {author} {\bibfnamefont {R.}~\bibnamefont {Nath}},
  \bibinfo {author} {\bibfnamefont {C.}~\bibnamefont {Geibel}}, \ and\ \bibinfo
  {author} {\bibfnamefont {H.}~\bibnamefont {Rosner}},\ }\href {\doibase
  10.1103/PhysRevB.77.104436} {\bibfield  {journal} {\bibinfo  {journal} {Phys.
  Rev. B}\ }\textbf {\bibinfo {volume} {77}},\ \bibinfo {pages} {104436}
  (\bibinfo {year} {2008})},\ \Eprint {http://arxiv.org/abs/arXiv:0802.2293}
  {arXiv:0802.2293} \BibitemShut {NoStop}%
\bibitem [{\citenamefont {Janson}\ \emph {et~al.}(2011)\citenamefont {Janson},
  \citenamefont {Tsirlin}, \citenamefont {Sichelschmidt}, \citenamefont
  {Skourski}, \citenamefont {Weickert},\ and\ \citenamefont
  {Rosner}}]{janson2011}%
  \BibitemOpen
  \bibfield  {author} {\bibinfo {author} {\bibfnamefont {O.}~\bibnamefont
  {Janson}}, \bibinfo {author} {\bibfnamefont {A.~A.}\ \bibnamefont {Tsirlin}},
  \bibinfo {author} {\bibfnamefont {J.}~\bibnamefont {Sichelschmidt}}, \bibinfo
  {author} {\bibfnamefont {Y.}~\bibnamefont {Skourski}}, \bibinfo {author}
  {\bibfnamefont {F.}~\bibnamefont {Weickert}}, \ and\ \bibinfo {author}
  {\bibfnamefont {H.}~\bibnamefont {Rosner}},\ }\href {\doibase
  10.1103/PhysRevB.83.094435} {\bibfield  {journal} {\bibinfo  {journal} {Phys.
  Rev. B}\ }\textbf {\bibinfo {volume} {83}},\ \bibinfo {pages} {094435}
  (\bibinfo {year} {2011})},\ \Eprint {http://arxiv.org/abs/arXiv:1011.5393}
  {arXiv:1011.5393} \BibitemShut {NoStop}%
\bibitem [{\citenamefont {Lebernegg}\ \emph {et~al.}(2011)\citenamefont
  {Lebernegg}, \citenamefont {Tsirlin}, \citenamefont {Janson}, \citenamefont
  {Nath}, \citenamefont {Sichelschmidt}, \citenamefont {Skourski},
  \citenamefont {Amthauer},\ and\ \citenamefont {Rosner}}]{lebernegg2011}%
  \BibitemOpen
  \bibfield  {author} {\bibinfo {author} {\bibfnamefont {S.}~\bibnamefont
  {Lebernegg}}, \bibinfo {author} {\bibfnamefont {A.~A.}\ \bibnamefont
  {Tsirlin}}, \bibinfo {author} {\bibfnamefont {O.}~\bibnamefont {Janson}},
  \bibinfo {author} {\bibfnamefont {R.}~\bibnamefont {Nath}}, \bibinfo {author}
  {\bibfnamefont {J.}~\bibnamefont {Sichelschmidt}}, \bibinfo {author}
  {\bibfnamefont {Y.}~\bibnamefont {Skourski}}, \bibinfo {author}
  {\bibfnamefont {G.}~\bibnamefont {Amthauer}}, \ and\ \bibinfo {author}
  {\bibfnamefont {H.}~\bibnamefont {Rosner}},\ }\href {\doibase
  10.1103/PhysRevB.84.174436} {\bibfield  {journal} {\bibinfo  {journal} {Phys.
  Rev. B}\ }\textbf {\bibinfo {volume} {84}},\ \bibinfo {pages} {174436}
  (\bibinfo {year} {2011})},\ \Eprint {http://arxiv.org/abs/arXiv:1107.0250}
  {arXiv:1107.0250} \BibitemShut {NoStop}%
\bibitem [{Note3()}]{Note3}%
  \BibitemOpen
  \bibinfo {note} {Here, we do not adapt the local coordinate frame to the
  global symmetry of the crystal structure, and direct the local $z$ axis along
  one of the Cr--O bonds (see also Fig.~\ref {F_wf}). Therefore, the resulting
  $d$ states do not show the weak splitting in the $t_{2g}$ subspace, as
  expected for trigonally distorted octahedron. This simplification does not
  affect any of our results, because Cr$^{3+}$ shows the robust high-spin state
  with the half-filled $t_{2g}$ levels, whereas fine structure of these levels
  has minor effect on the magnetism.}\BibitemShut {Stop}%
\bibitem [{\citenamefont {Johnston}\ \emph {et~al.}(2000)\citenamefont
  {Johnston}, \citenamefont {Kremer}, \citenamefont {Troyer}, \citenamefont
  {Wang}, \citenamefont {Kl{\"{u}}mper}, \citenamefont {Bud'ko}, \citenamefont
  {Panchula},\ and\ \citenamefont {Canfield}}]{HC_AHC_Johnston}%
  \BibitemOpen
  \bibfield  {author} {\bibinfo {author} {\bibfnamefont {D.~C.}\ \bibnamefont
  {Johnston}}, \bibinfo {author} {\bibfnamefont {R.~K.}\ \bibnamefont
  {Kremer}}, \bibinfo {author} {\bibfnamefont {M.}~\bibnamefont {Troyer}},
  \bibinfo {author} {\bibfnamefont {X.}~\bibnamefont {Wang}}, \bibinfo {author}
  {\bibfnamefont {A.}~\bibnamefont {Kl{\"{u}}mper}}, \bibinfo {author}
  {\bibfnamefont {S.~L.}\ \bibnamefont {Bud'ko}}, \bibinfo {author}
  {\bibfnamefont {A.~F.}\ \bibnamefont {Panchula}}, \ and\ \bibinfo {author}
  {\bibfnamefont {P.~C.}\ \bibnamefont {Canfield}},\ }\href {\doibase
  10.1103/PhysRevB.61.9558} {\bibfield  {journal} {\bibinfo  {journal} {Phys.
  Rev. B}\ }\textbf {\bibinfo {volume} {61}},\ \bibinfo {pages} {9558}
  (\bibinfo {year} {2000})},\ \Eprint {http://arxiv.org/abs/cond-mat/0003271}
  {cond-mat/0003271} \BibitemShut {NoStop}%
\bibitem [{\citenamefont {Czy{\.z}yk}\ and\ \citenamefont
  {Sawatzky}(1994)}]{LDA_U_AMF_FLL}%
  \BibitemOpen
  \bibfield  {author} {\bibinfo {author} {\bibfnamefont {M.~T.}\ \bibnamefont
  {Czy{\.z}yk}}\ and\ \bibinfo {author} {\bibfnamefont {G.~A.}\ \bibnamefont
  {Sawatzky}},\ }\href {\doibase 10.1103/PhysRevB.49.14211} {\bibfield
  {journal} {\bibinfo  {journal} {Phys. Rev. B}\ }\textbf {\bibinfo {volume}
  {49}},\ \bibinfo {pages} {14211} (\bibinfo {year} {1994})}\BibitemShut
  {NoStop}%
\bibitem [{\citenamefont {Mazin}(2007)}]{LiCrO2_DFT_2007}%
  \BibitemOpen
  \bibfield  {author} {\bibinfo {author} {\bibfnamefont {I.~I.}\ \bibnamefont
  {Mazin}},\ }\href {\doibase 10.1103/PhysRevB.75.094407} {\bibfield  {journal}
  {\bibinfo  {journal} {Phys. Rev. B}\ }\textbf {\bibinfo {volume} {75}},\
  \bibinfo {pages} {094407} (\bibinfo {year} {2007})},\ \Eprint
  {http://arxiv.org/abs/arXiv:cond-mat/0701520} {arXiv:cond-mat/0701520}
  \BibitemShut {NoStop}%
\bibitem [{\citenamefont {Fennie}\ and\ \citenamefont
  {Rabe}(2006)}]{fennie2006}%
  \BibitemOpen
  \bibfield  {author} {\bibinfo {author} {\bibfnamefont {C.~J.}\ \bibnamefont
  {Fennie}}\ and\ \bibinfo {author} {\bibfnamefont {K.~M.}\ \bibnamefont
  {Rabe}},\ }\href {\doibase 10.1103/PhysRevLett.96.205505} {\bibfield
  {journal} {\bibinfo  {journal} {Phys. Rev. Lett.}\ }\textbf {\bibinfo
  {volume} {96}},\ \bibinfo {pages} {205505} (\bibinfo {year} {2006})},\
  \Eprint {http://arxiv.org/abs/cond-mat/0602503} {cond-mat/0602503}
  \BibitemShut {NoStop}%
\bibitem [{\citenamefont {Franchini}\ \emph {et~al.}(2007)\citenamefont
  {Franchini}, \citenamefont {Podloucky}, \citenamefont {Paier}, \citenamefont
  {Marsman},\ and\ \citenamefont {Kresse}}]{franchini2007}%
  \BibitemOpen
  \bibfield  {author} {\bibinfo {author} {\bibfnamefont {C.}~\bibnamefont
  {Franchini}}, \bibinfo {author} {\bibfnamefont {R.}~\bibnamefont
  {Podloucky}}, \bibinfo {author} {\bibfnamefont {J.}~\bibnamefont {Paier}},
  \bibinfo {author} {\bibfnamefont {M.}~\bibnamefont {Marsman}}, \ and\
  \bibinfo {author} {\bibfnamefont {G.}~\bibnamefont {Kresse}},\ }\href
  {\doibase 10.1103/PhysRevB.75.195128} {\bibfield  {journal} {\bibinfo
  {journal} {Phys. Rev. B}\ }\textbf {\bibinfo {volume} {75}},\ \bibinfo
  {pages} {195128} (\bibinfo {year} {2007})}\BibitemShut {NoStop}%
\bibitem [{\citenamefont {Iori}\ \emph {et~al.}(2012)\citenamefont {Iori},
  \citenamefont {Gatti},\ and\ \citenamefont {Rubio}}]{iori2012}%
  \BibitemOpen
  \bibfield  {author} {\bibinfo {author} {\bibfnamefont {F.}~\bibnamefont
  {Iori}}, \bibinfo {author} {\bibfnamefont {M.}~\bibnamefont {Gatti}}, \ and\
  \bibinfo {author} {\bibfnamefont {A.}~\bibnamefont {Rubio}},\ }\href
  {\doibase 10.1103/PhysRevB.85.115129} {\bibfield  {journal} {\bibinfo
  {journal} {Phys. Rev. B}\ }\textbf {\bibinfo {volume} {85}},\ \bibinfo
  {pages} {115129} (\bibinfo {year} {2012})},\ \Eprint
  {http://arxiv.org/abs/arXiv:1201.3308} {arXiv:1201.3308} \BibitemShut
  {NoStop}%
\bibitem [{\citenamefont {Tsirlin}\ \emph {et~al.}(2010)\citenamefont
  {Tsirlin}, \citenamefont {Janson},\ and\ \citenamefont
  {Rosner}}]{tsirlin2010}%
  \BibitemOpen
  \bibfield  {author} {\bibinfo {author} {\bibfnamefont {A.~A.}\ \bibnamefont
  {Tsirlin}}, \bibinfo {author} {\bibfnamefont {O.}~\bibnamefont {Janson}}, \
  and\ \bibinfo {author} {\bibfnamefont {H.}~\bibnamefont {Rosner}},\ }\href
  {\doibase 10.1103/PhysRevB.82.144416} {\bibfield  {journal} {\bibinfo
  {journal} {Phys. Rev. B}\ }\textbf {\bibinfo {volume} {82}},\ \bibinfo
  {pages} {144416} (\bibinfo {year} {2010})},\ \Eprint
  {http://arxiv.org/abs/arXiv:1007.1646} {arXiv:1007.1646} \BibitemShut
  {NoStop}%
\bibitem [{\citenamefont {Tsirlin}\ \emph
  {et~al.}(2011{\natexlab{a}})\citenamefont {Tsirlin}, \citenamefont {Janson},\
  and\ \citenamefont {Rosner}}]{tsirlin2011}%
  \BibitemOpen
  \bibfield  {author} {\bibinfo {author} {\bibfnamefont {A.~A.}\ \bibnamefont
  {Tsirlin}}, \bibinfo {author} {\bibfnamefont {O.}~\bibnamefont {Janson}}, \
  and\ \bibinfo {author} {\bibfnamefont {H.}~\bibnamefont {Rosner}},\ }\href
  {\doibase 10.1103/PhysRevB.84.144429} {\bibfield  {journal} {\bibinfo
  {journal} {Phys. Rev. B}\ }\textbf {\bibinfo {volume} {84}},\ \bibinfo
  {pages} {144429} (\bibinfo {year} {2011}{\natexlab{a}})},\ \Eprint
  {http://arxiv.org/abs/arXiv:1104.2495} {arXiv:1104.2495} \BibitemShut
  {NoStop}%
\bibitem [{Note4()}]{Note4}%
  \BibitemOpen
  \bibinfo {note} {For the fitting, we used the experimental data above
  $T_N$\protect \tmspace +\thinmuskip {.1667em}=\protect \tmspace +\thinmuskip
  {.1667em}28\protect \tmspace +\thinmuskip {.1667em}K, since a one-dimensional
  model cannot account for the AF ordering (Ref.~\protect \rev@citealpnum
  {Mermin_Wagner}).}\BibitemShut {Stop}%
\bibitem [{\citenamefont {Mermin}\ and\ \citenamefont
  {Wagner}(1966)}]{Mermin_Wagner}%
  \BibitemOpen
  \bibfield  {author} {\bibinfo {author} {\bibfnamefont {N.~D.}\ \bibnamefont
  {Mermin}}\ and\ \bibinfo {author} {\bibfnamefont {H.}~\bibnamefont
  {Wagner}},\ }\href {\doibase 10.1103/PhysRevLett.17.1133} {\bibfield
  {journal} {\bibinfo  {journal} {Phys. Rev. Lett.}\ }\textbf {\bibinfo
  {volume} {17}},\ \bibinfo {pages} {1133} (\bibinfo {year}
  {1966})}\BibitemShut {NoStop}%
\bibitem [{Note5()}]{Note5}%
  \BibitemOpen
  \bibinfo {note} {Unfortunately, the quasi-1D character of the magnetic model
  requires $L_z\gg {}L_x\simeq {}L_y$ for an $N$-site lattice ($N\protect
  \tmspace -\thinmuskip {.1667em}=\protect \tmspace -\thinmuskip
  {.1667em}L_x\times {}L_y\times {}L_z$). This leads to a small number of
  computationally feasible lattices, impeding an accurate estimation of the
  magnetic moment. To check the results for consistency, we repeated the
  fitting using a simplified scaling $\sigma _2\protect \tmspace -\thinmuskip
  {.1667em}=\protect \tmspace -\thinmuskip {.1667em}0$. As expected, the
  $\sigma _1$ values are substantially renormalized. Yet, both approaches yield
  marginally different values of $S_{\infty }$ ($\sim $1\protect \tmspace
  +\thinmuskip {.1667em}\% difference).}\BibitemShut {Stop}%
\bibitem [{\citenamefont {Yajima}\ and\ \citenamefont
  {Takahashi}(1996)}]{yajima1996}%
  \BibitemOpen
  \bibfield  {author} {\bibinfo {author} {\bibfnamefont {M.}~\bibnamefont
  {Yajima}}\ and\ \bibinfo {author} {\bibfnamefont {M.}~\bibnamefont
  {Takahashi}},\ }\href {\doibase 10.1143/JPSJ.65.39} {\bibfield  {journal}
  {\bibinfo  {journal} {J. Phys. Soc. Jpn.}\ }\textbf {\bibinfo {volume}
  {65}},\ \bibinfo {pages} {39} (\bibinfo {year} {1996})}\BibitemShut {NoStop}%
\bibitem [{\citenamefont {Yamamoto}(1997)}]{yamamoto1997}%
  \BibitemOpen
  \bibfield  {author} {\bibinfo {author} {\bibfnamefont {S.}~\bibnamefont
  {Yamamoto}},\ }\href {\doibase 10.1103/PhysRevB.55.3603} {\bibfield
  {journal} {\bibinfo  {journal} {Phys. Rev. B}\ }\textbf {\bibinfo {volume}
  {55}},\ \bibinfo {pages} {3603} (\bibinfo {year} {1997})}\BibitemShut
  {NoStop}%
\bibitem [{\citenamefont {Affleck}\ \emph {et~al.}(1987)\citenamefont
  {Affleck}, \citenamefont {Kennedy}, \citenamefont {Lieb},\ and\ \citenamefont
  {Tasaki}}]{affleck1987}%
  \BibitemOpen
  \bibfield  {author} {\bibinfo {author} {\bibfnamefont {I.}~\bibnamefont
  {Affleck}}, \bibinfo {author} {\bibfnamefont {T.}~\bibnamefont {Kennedy}},
  \bibinfo {author} {\bibfnamefont {E.~H.}\ \bibnamefont {Lieb}}, \ and\
  \bibinfo {author} {\bibfnamefont {H.}~\bibnamefont {Tasaki}},\ }\href
  {\doibase 10.1103/PhysRevLett.59.799} {\bibfield  {journal} {\bibinfo
  {journal} {Phys. Rev. Lett.}\ }\textbf {\bibinfo {volume} {59}},\ \bibinfo
  {pages} {799} (\bibinfo {year} {1987})}\BibitemShut {NoStop}%
\bibitem [{\citenamefont {Affleck}\ \emph {et~al.}(1988)\citenamefont
  {Affleck}, \citenamefont {Kennedy}, \citenamefont {Lieb},\ and\ \citenamefont
  {Tasaki}}]{affleck1988}%
  \BibitemOpen
  \bibfield  {author} {\bibinfo {author} {\bibfnamefont {I.}~\bibnamefont
  {Affleck}}, \bibinfo {author} {\bibfnamefont {T.}~\bibnamefont {Kennedy}},
  \bibinfo {author} {\bibfnamefont {E.~H.}\ \bibnamefont {Lieb}}, \ and\
  \bibinfo {author} {\bibfnamefont {H.}~\bibnamefont {Tasaki}},\ }\href
  {\doibase 10.1007/BF01218021} {\bibfield  {journal} {\bibinfo  {journal}
  {Comm. Math. Phys.}\ }\textbf {\bibinfo {volume} {115}},\ \bibinfo {pages}
  {477} (\bibinfo {year} {1988})}\BibitemShut {NoStop}%
\bibitem [{\citenamefont {Tennant}\ \emph {et~al.}(1997)\citenamefont
  {Tennant}, \citenamefont {Nagler}, \citenamefont {Garrett}, \citenamefont
  {Barnes},\ and\ \citenamefont {Torardi}}]{tennant1997}%
  \BibitemOpen
  \bibfield  {author} {\bibinfo {author} {\bibfnamefont {D.~A.}\ \bibnamefont
  {Tennant}}, \bibinfo {author} {\bibfnamefont {S.~E.}\ \bibnamefont {Nagler}},
  \bibinfo {author} {\bibfnamefont {A.~W.}\ \bibnamefont {Garrett}}, \bibinfo
  {author} {\bibfnamefont {T.}~\bibnamefont {Barnes}}, \ and\ \bibinfo {author}
  {\bibfnamefont {C.~C.}\ \bibnamefont {Torardi}},\ }\href {\doibase
  10.1103/PhysRevLett.78.4998} {\bibfield  {journal} {\bibinfo  {journal}
  {Phys. Rev. Lett.}\ }\textbf {\bibinfo {volume} {78}},\ \bibinfo {pages}
  {4998} (\bibinfo {year} {1997})},\ \Eprint
  {http://arxiv.org/abs/cond-mat/9704093} {cond-mat/9704093} \BibitemShut
  {NoStop}%
\bibitem [{Note6()}]{Note6}%
  \BibitemOpen
  \bibinfo {note} {The interchain couplings should be considered irrespective
  of their sign and together with the relevant coordination numbers. In
  Cr$_2[$BP$_3$O$_{12}]$, the effective interchain coupling is $J_{\protect
  \text {eff}}=3J_{\protect \rm ic2}+6|J_{\protect \rm ic1}|=0.18J_1$, i.e.,
  $J_{\protect \rm ic}/J_1=0.09$ for the typical coordination number of 2 as,
  e.g., in Pb$_2[$V$_3$O$_9]$.}\BibitemShut {Stop}%
\bibitem [{\citenamefont {Tsirlin}\ and\ \citenamefont
  {Rosner}(2011)}]{tsirlin2011c}%
  \BibitemOpen
  \bibfield  {author} {\bibinfo {author} {\bibfnamefont {A.~A.}\ \bibnamefont
  {Tsirlin}}\ and\ \bibinfo {author} {\bibfnamefont {H.}~\bibnamefont
  {Rosner}},\ }\href {\doibase 10.1103/PhysRevB.83.064415} {\bibfield
  {journal} {\bibinfo  {journal} {Phys. Rev. B}\ }\textbf {\bibinfo {volume}
  {83}},\ \bibinfo {pages} {064415} (\bibinfo {year} {2011})},\ \Eprint
  {http://arxiv.org/abs/arXiv:1011.3981} {arXiv:1011.3981} \BibitemShut
  {NoStop}%
\bibitem [{\citenamefont {Tsirlin}\ \emph
  {et~al.}(2011{\natexlab{b}})\citenamefont {Tsirlin}, \citenamefont {Nath},
  \citenamefont {Sichelschmidt}, \citenamefont {Skourski}, \citenamefont
  {Geibel},\ and\ \citenamefont {Rosner}}]{tsirlin2011b}%
  \BibitemOpen
  \bibfield  {author} {\bibinfo {author} {\bibfnamefont {A.~A.}\ \bibnamefont
  {Tsirlin}}, \bibinfo {author} {\bibfnamefont {R.}~\bibnamefont {Nath}},
  \bibinfo {author} {\bibfnamefont {J.}~\bibnamefont {Sichelschmidt}}, \bibinfo
  {author} {\bibfnamefont {Y.}~\bibnamefont {Skourski}}, \bibinfo {author}
  {\bibfnamefont {C.}~\bibnamefont {Geibel}}, \ and\ \bibinfo {author}
  {\bibfnamefont {H.}~\bibnamefont {Rosner}},\ }\href {\doibase
  10.1103/PhysRevB.83.144412} {\bibfield  {journal} {\bibinfo  {journal} {Phys.
  Rev. B}\ }\textbf {\bibinfo {volume} {83}},\ \bibinfo {pages} {144412}
  (\bibinfo {year} {2011}{\natexlab{b}})},\ \Eprint
  {http://arxiv.org/abs/arXiv:1101.2546} {arXiv:1101.2546} \BibitemShut
  {NoStop}%
\end{thebibliography}

%

\setcounter{figure}{0}
\setcounter{table}{0}
\renewcommand{\thefigure}{S\arabic{figure}}
\renewcommand{\thetable}{S\arabic{table}}
\begin{widetext}
\vskip 12cm
\begin{center}
{\large
Supplementary information for 
\smallskip

\textbf{Structure and magnetism of Cr$_2[$BP$_3$O$_{12}]$:\\ Towards the ``quantum--classical'' crossover in a spin-3/2 alternating chain}}
\medskip

O. Janson, S. Chen, A. A. Tsirlin, S. Hoffmann, J. Sichelschmidt,

Q. Huang, Z.-J. Zhang, M.-B. Tang, J.-T. Zhao, R. Kniep, H. Rosner
\end{center}
\medskip

\begin{figure}[!h]
\includegraphics[width=11cm]{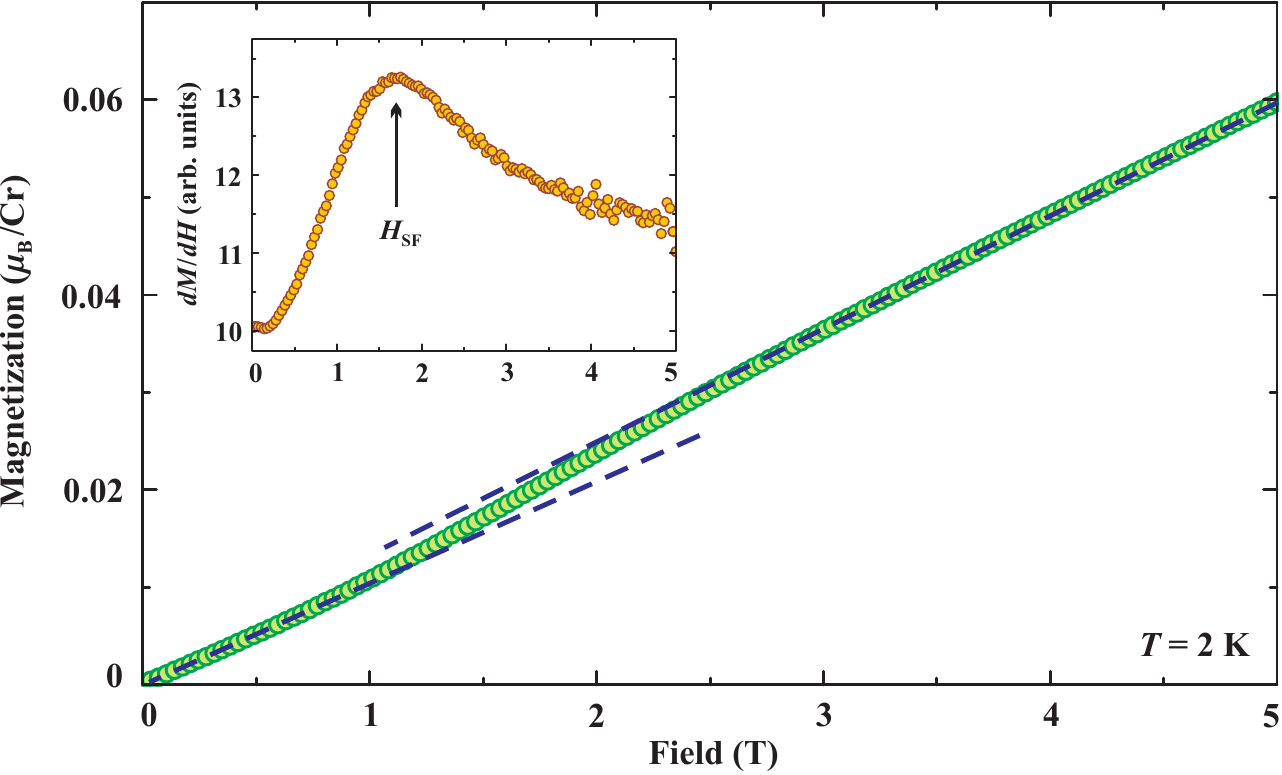}
\begin{minipage}{14cm}
\caption{\label{fig:s1}\normalsize
Field dependence of the magnetization of Cr$_2[$BP$_3$O$_{12}]$ measured at 2~K.
The lines are guide-for-the-eye. The inset shows the field derivative of the
magnetization. 
}
\end{minipage}
\end{figure}
\bigskip

\begin{figure}[!h]
\includegraphics[width=11cm]{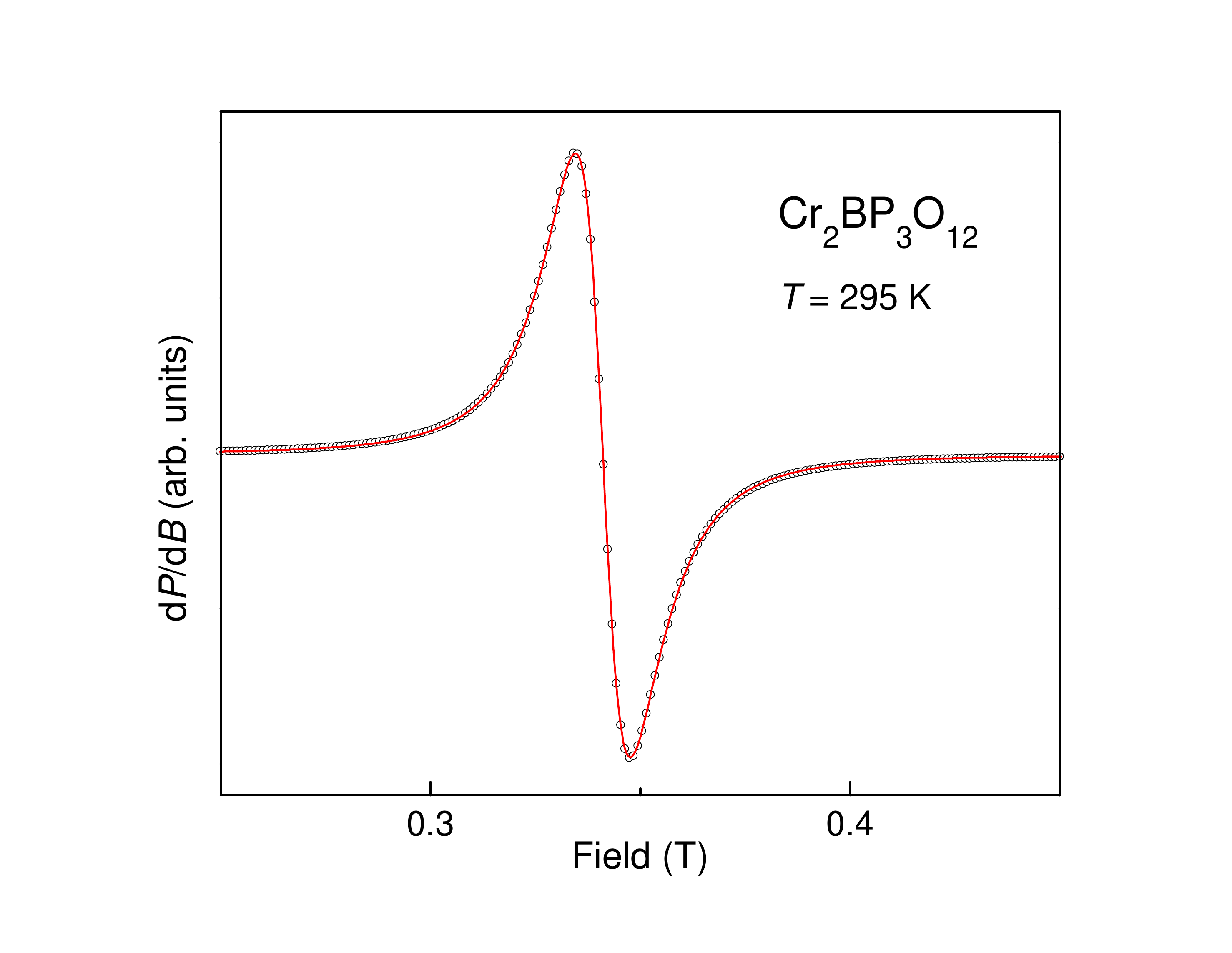}
\begin{minipage}{14cm}
\caption{\label{fig:s2}\normalsize
Electron spin resonance spectrum (circles; first derivative of absorbed
microwave power $dP/dB$) taken at 9.5\,GHz (X-band). Solid line denotes a
powder-averaged Lorentzian line with $g_\perp\approx g_\| \!=\!1.9680\pm0.0005$
and linewidth $\Delta B_\perp\!=\!8.6\pm0.2$~mT and $\Delta
B_\|\!=\!18.8\pm0.2$\,mT.
}
\end{minipage}
\end{figure}
\bigskip

\begin{figure}[!h]
\includegraphics[width=11cm]{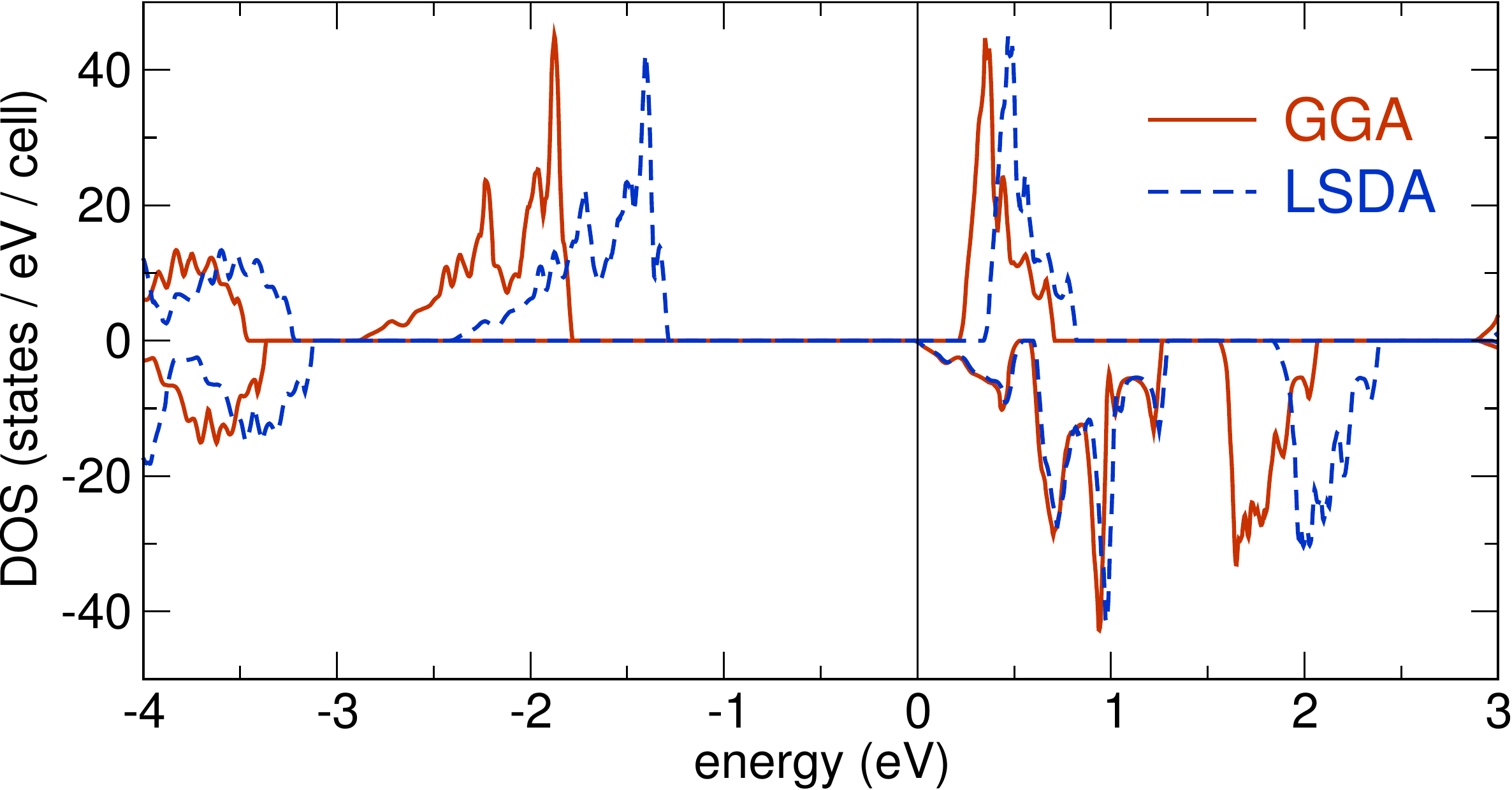}
\begin{minipage}{14cm}
\caption{\label{fig:s3}\normalsize
Cr$_2[$BP$_3$O$_{12}]$: GGA (solid line) and LSDA (dashed line) density of
states for the ferromagnetic solution. Positive and negative values refer to
the majority and minority spin channel, respectively. Note the different values
of the band gap: $-1.78$\,eV in GGA and $-1.29$\,eV in LSDA.}
\end{minipage}
\end{figure}
\bigskip

\begin{figure}[!h]
\includegraphics[width=11cm]{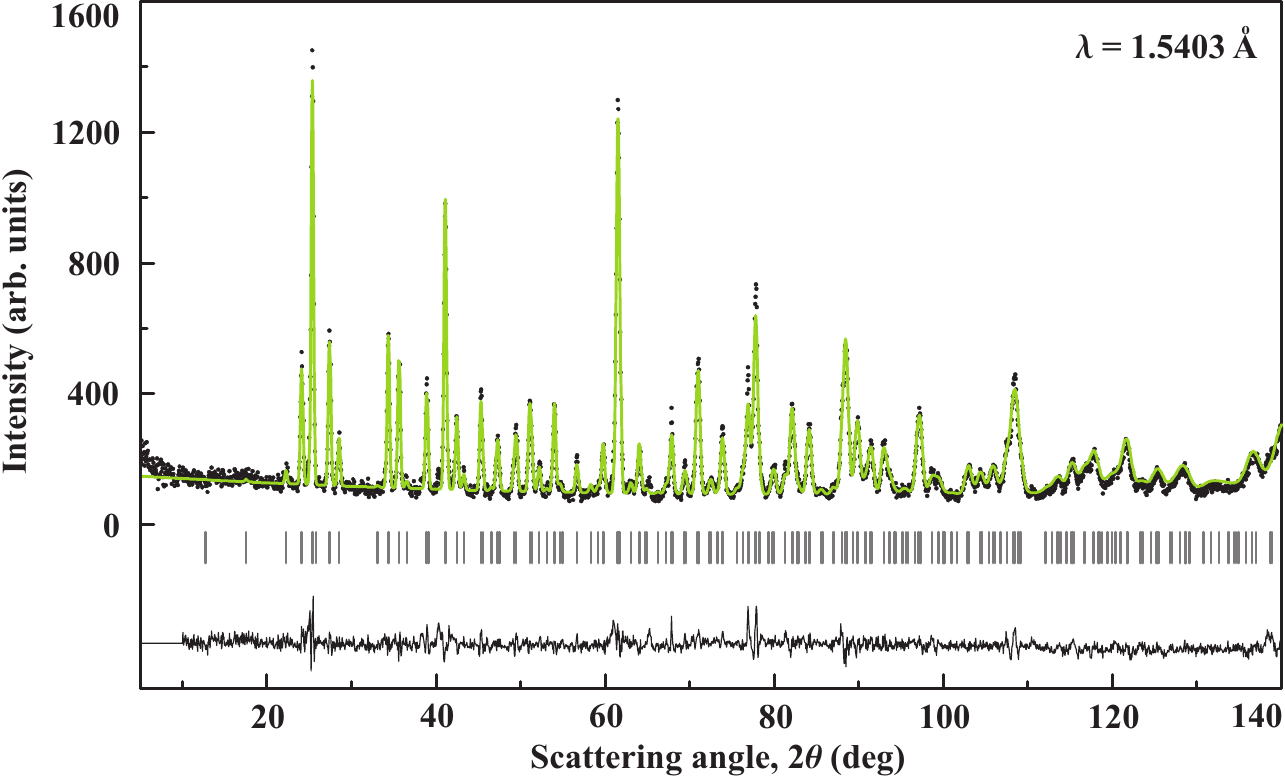}
\begin{minipage}{14cm}
\caption{\label{fig:s4}\normalsize
Rietveld refinement of the neutron diffraction data collected at 35~K. Ticks show the reflection positions.}
\end{minipage}
\end{figure}

\end{widetext}
\end{document}